\documentclass[useAMS,usenatbib]{mn2e}
\usepackage{natbib}
\usepackage{deluxetable}
\usepackage{threeparttable}
\usepackage{longtable}
\usepackage{environ}
\usepackage{graphicx}
\usepackage{textcomp}
\usepackage[fleqn]{amsmath}
\usepackage{lscape}
\usepackage{pdflscape}
\usepackage{dcolumn}
\usepackage{rccol}
\usepackage{threeparttablex}
\usepackage{amssymb}
\usepackage{comment} 

\newcommand{\about}{$\sim\!\!$~}
\newcommand{\bvri}{\protect\hbox{$BV\!RI$} }
\newcommand{\ugriz}{\protect\hbox{$ugriz$} }
\newcommand{\rizyjh}{\protect\hbox{$riZY\!\!J\!H$} }
\newcommand{\yjh}{\protect\hbox{$Y\!\!J\!H$} }
\newcommand{\bvrriizyjh}{\protect\hbox{$BV\!rRiIZY\!\!J\!H$} }
\newcommand{\kms}{\,km\,s$^{-1}$}

\newcommand{\mstellar}{\ensuremath{M_{\mathrm{stellar}}}}
\newcommand{\omatter}{\ensuremath{\Omega_{\mathrm{M}}}}

\newcommand{\absm}{\ensuremath{M_B}}

\newcommand{\deltam}{\ensuremath{\Delta m_{15}}}

\newcommand{\oi}{\ensuremath{\mathrm{O}\,\textsc{i}}}

\newcommand{\cii}{\ensuremath{\mathrm{C}\,\textsc{ii}\,\lambda6580}}
\newcommand{\ciitmp}{\ensuremath{\mathrm{C}\,\textsc{ii}\,\lambda7234}}

\newcommand{\Caii}{\ensuremath{\mathrm{Ca}\,\textsc{ii}}}

\newcommand{\Siii}{\ensuremath{\mathrm{Si}\,\textsc{ii}\,\lambda6355}}

\newcommand{\Siiitmp}{\ensuremath{\mathrm{Si}\,\textsc{ii}\,\lambda5972}}
\newcommand{\Siiitmpsec}{\ensuremath{\mathrm{Si}\,\textsc{ii}\,\lambda4130}}
\newcommand{\Feii}{\ensuremath{\mathrm{Fe}\,\textsc{ii}\,\lambda3250}}
\newcommand{\Feiineb}{\ensuremath{[\mathrm{Fe}\,\textsc{ii}]\,\lambda7155}}
\newcommand{\Feiii}{\ensuremath{[\mathrm{Fe}\,\textsc{iii}]\,\lambda4701}}
\newcommand{\Coiii}{\ensuremath{[\mathrm{Co}\,\textsc{iii}]\,\lambda5891}}
\newcommand{\Niii}{\ensuremath{[\mathrm{Ni}\,\textsc{ii}]\,\lambda7378}}

\newcommand{\Nifs}{$^{56}$Ni}
\newcommand{\Nife}{$^{58}$Ni}
\newcommand{\nai}{\ensuremath{\mathrm{Na}\,\textsc{i}}}

\title[Type Ia SN~2013dy]{500 Days of SN~2013dy: spectra and photometry from the ultraviolet to the infrared}
\author[Pan et al.]{
Y.-C. Pan$^{1}$\thanks{E-mail:ycpan@illinois.edu}, 
R. J. Foley$^{1,2}$, 
M. Kromer$^{3}$,
O. D. Fox$^{4}$,
W. Zheng$^{4}$,
P. Challis$^{5}$,\newauthor
K. I. Clubb$^{4}$,
A. V. Filippenko$^{4}$,
G. Folatelli$^{6}$,
M. L. Graham$^{4}$,
W. Hillebrandt$^{7}$,\newauthor
R. P. Kirshner$^{5}$,
W. H. Lee$^{8}$,
R. Pakmor$^{9}$, 
F. Patat$^{10}$,
M. M. Phillips$^{11}$,\newauthor 
G. Pignata$^{12,13}$,
F. R{\"o}pke$^{14,15}$,
I. Seitenzahl$^{16}$,
J. M. Silverman$^{17,18}$, 
J. D. Simon$^{19}$,\newauthor
A. Sternberg$^{20}$,
M. D. Stritzinger$^{21}$, 
S. Taubenberger$^{7}$,
J. Vinko$^{17,22}$, 
J. C. Wheeler$^{17}$\\
  $^{1}$Astronomy Department, University of Illinois at Urbana-Champaign, 1002 W.\ Green Street, Urbana, IL 61801, USA\\
  $^{2}$Department of Physics, University of Illinois Urbana-Champaign, 1110 W.\ Green Street, Urbana, IL 61801, USA\\
  $^{3}$The Oskar Klein Centre \& Department of Astronomy, Stockholm University, AlbaNova, SE-106 91 Stockholm, Sweden\\
  $^{4}$Department of Astronomy, University of California, Berkeley, CA 94720-3411, USA\\
  $^{5}$Harvard-Smithsonian Center for Astrophysics, 60 Garden Street, Cambridge, MA 02138, USA\\
  $^{6}$Kavli Institute for the Physics and Mathematics of the Universe (WPI), The University of Tokyo, Kashiwa, Chiba 277-8583, Japan\\
  $^{7}$Max-Planck Institut f{\"u}r Astrophysik, Karl-Schwarzschild-Str. 1, D-85748 Garching b. M\"unchen, Germany\\
  $^{8}$Instituto de Astronomia Universidad Nacional Autonoma de Mexico, Apdo. Postal 70-264, Cd. Universitaria, Mexico DF 04510, Mexico\\
  $^{9}$Heidelberger Institut f\"{u}r Theoretische Studien, Schloss-Wolfsbrunnenweg 35, 69118 Heidelberg, Germany\\
  $^{10}$European Organisation for Astronomical Research in the Southern Hemisphere (ESO), Karl-Schwarzschild-Str. 2, 85748 Garching b. \\
  Munchen, Germany\\ 
  $^{11}$Carnegie Observatories, Las Campanas Observatory, La Serena, Chile\\
  $^{12}$Departamento de Ciencias Fisicas, Universidad Andres Bello, Avda. Republica 252, Santiago, Chile\\
  $^{13}$Millennium Institute of Astrophysics, Chile\\
  $^{14}$Heidelberger Institut f{\"u}r Theoretische Studien, Schloss-Wolfsbrunnenweg 35, 69118 Heidelberg, Germany\\
  $^{15}$Zentrum f{\"u}r Astronomie der Universit{\"a}t Heidelberg, Institut f{\"u}r Theoretische Astrophysik,
  Philosophenweg 12, 69102 Heidelberg, Germany\\
  $^{16}$Research School of Astronomy and Astrophysics, Australian National University, Cotter Road, Weston Creek, ACT, 2611, Australia\\
  $^{17}$Department of Astronomy, University of Texas at Austin, Austin, TX 78712, USA\\
  $^{18}$NSF Astronomy and Astrophysics Postdoctoral Fellow\\
  $^{19}$Observatories of the Carnegie Institution for Science, 813 Santa Barbara St., Pasadena, CA 91101, USA\\
  $^{20}$Excellence Cluster Universe, Technische UniversitŠt MŸnchen, Boltzmannstr. 2, D-85748, Garching, Germany\\
  $^{21}$Department of Physics and Astronomy, Aarhus University, Ny Munkegade, DK-8000 Aarhus C, Denmark\\
  $^{22}$Department of Optics and Quantum Electronics, University of Szeged, Szeged, Dem ter 9, 6720, Hungary\\
}

\begin{document}

\maketitle

\label{firstpage}

\begin{abstract}
  SN~2013dy is a Type Ia supernova for which we have compiled an
  extraordinary dataset spanning from 0.1 to \about 500~days after
  explosion.  We present 10 epochs of ultraviolet (UV) through
  near-infrared (NIR) spectra with \textit{HST}/STIS, 47 epochs
  of optical spectra (15 of them having high resolution), and
  more than 500 photometric observations in the \bvrriizyjh\ bands.
  SN~2013dy has a broad and slowly declining light curve
  ($\deltam(B) = 0.92$~mag), shallow \Siii\ absorption, and a low
  velocity gradient.  We detect strong \ion{C}{2} in our earliest
  spectra, probing unburned progenitor material in the outermost
  layers of the SN ejecta, but this feature fades within a few days.
  The UV continuum of SN~2013dy, which is strongly affected by the
  metal abundance of the progenitor star, suggests that SN~2013dy had
  a relatively high-metallicity progenitor.  Examining one of the largest
  single set of high-resolution spectra for a SN~Ia, we find no
  evidence of variable absorption from circumstellar material.
  Combining our UV spectra, NIR photometry, and high-cadence optical
  photometry, we construct a bolometric light curve, showing that
  SN~2013dy had a maximum luminosity of $10.0^{+4.8}_{-3.8} \times
  10^{42}$~erg~s$^{-1}$.  We compare the synthetic light curves and
  spectra of several models to SN~2013dy, finding that SN~2013dy is
  in good agreement with a solar-metallicity W7 model.
\end{abstract}


\section{Introduction}
\label{sec:introduction}
Type Ia supernovae (SNe~Ia) are important distance indicators.
Observations of SNe~Ia provided the first evidence of the accelerating
expansion of the universe \citep{1998AJ....116.1009R,
  1999ApJ...517..565P}.  Observational evidence indicates that they
are the result of the thermonuclear explosion of an accreting
carbon-oxygen white dwarf (WD) star in a close binary system
\citep[e.g.,][]{2000ARA&A..38..191H, 2013FrPhy...8..116H,
2014ARA&A..52..107M}.  Although recent observations have constrained
the exploding star to be consistent with a compact object
\citep{2011Natur.480..344N, 2012ApJ...744L..17B}, the nature of the
presumed companion star is not yet clear. 

Some explosion mechanisms have been proposed to explain the observations of SNe Ia.
For a review of explosion models, see \citet{2000ARA&A..38..191H}
and \citet{2013FrPhy...8..116H}. A pure deflagration model 
that results from a subsonic deflagration flame can produce 
intermediate-mass elements (IME), but it fails to synthesize enough 
iron-group elements (IGEs) to meet the constraints of 
normal SNe Ia. However, a pure deflagration may explain some
subclasses of thermonuclear explosions 
\citep{2007PASP..119..360P,2013ApJ...767...57F,2013MNRAS.429.2287K,2014MNRAS.438.1762F}.
To create more IGEs, a model in which a supersonic detonation
following the deflagration (the so-called delayed-detonation model) is proposed.
This model produces more IGEs and better explains normal
SNe~Ia than the pure deflagration model \citep[e.g.,][]{2013MNRAS.436..333S}.

Statistical studies using a large sample of SNe~Ia can be useful in
understanding the physical properties of SN~Ia explosions and
progenitor systems.  However, a well-studied single SN~Ia with
complete and high-quality observations can provide highly
constraining information as well.  Recent SNe~Ia with extensive datasets
include SNe~2009ig \citep{2012ApJ...744...38F,2013ApJ...777...40M}, 2011fe
\citep[e.g.,][]{2011Natur.480..344N, 2011Natur.480..348L,
2012ApJ...744L..17B}, 2012cg \citep{2012ApJ...756L...7S}, 
2012fr \citep{2013ApJ...770...29C}, and
2014J \citep[e.g.,][]{2014ApJ...783L..24Z, 2014ApJ...784L..12G,
2014ApJ...788L..21A, 2014MNRAS.443.2887F,2015ApJ...798...39M}.
These SNe~Ia are all
very close ($D \leq 20$~Mpc), which allows for comprehensive datasets
including observations at non-optical wavelengths and extremely
late-time data, and discovered soon after explosion, which provides
information about the outermost layers of the ejecta.  For
each of these nearby SNe~Ia, large follow-up campaigns were initiated.
These high-quality datasets have resulted in some of the most
important constraints for the progenitor systems and explosions of SNe~Ia
\citep[e.g., progenitor metallicity;][]{2014MNRAS.439.1959M}.

One of the most critical pieces of information one can obtain for a
SN~Ia is a series of ultraviolet (UV) spectra.  Theoretical studies
show that the metallicity of the progenitor can significantly impact
both the photometric and spectroscopic properties of SNe~Ia
\citep{1998ApJ...495..617H, 2000ApJ...530..966L, 2003ApJ...590L..83T,
  2009Natur.460..869K, 2012MNRAS.427..103W}, but has little effect on optical 
and near-infrared (NIR) spectral properties.  While progenitor metallicity
does not affect the shape of the optical spectral-energy distribution
(SED), the optical spectral features, or the light-curve shape
\citep{1998ApJ...495..617H, 2000ApJ...530..966L, 2006astro.ph..8324P}, 
it is expected to affect the amount of $^{56}$Ni generated in the explosion
and thus the peak luminosity \citep{2003ApJ...590L..83T}.  A different peak
luminosity for SNe~Ia with the same light-curve shape could introduce
a large systematic bias in cosmological distance estimates, especially
if the average progenitor metallicity evolves with redshift.

While observations of SN environments indicate that metallicity could
subtly affect observables \citep{2005ApJ...634..210G,
  2011ApJ...743..172D, 2012A&A...545A..58S, 2013MNRAS.435.1680J,
  2013ApJ...770..108C, 2014MNRAS.438.1391P}, the most powerful and
indicative constraints have come from analyses of UV spectra.
\citet{2013ApJ...769L...1F} examined the UV spectra of the ``twin''
SNe~2011by and 2011fe, which had nearly identical optical SEDs,
optical colors, and optical light-curve shapes.  However, their UV
spectra differed significantly, such that the progenitor of SN~2011by
had a higher metallicity than that of SN~2011fe.  Intriguingly, their
peak luminosities differed in a way that one would predict based on
the progenitor metallicity differences; however, some or all of this
difference may be caused by an incorrect distance to SN~2011by
\citep{2015MNRAS.446.2073G}.  Regardless of the distance to SN~2011by, this
represents the first robust detection of different metallicities for
SN~Ia progenitors.  Additional direct analyses of UV spectral time
series have placed constraints on SN~Ia progenitor metallicity
\citep{2013MNRAS.429.2228H, 2014MNRAS.439.1959M}.

In this work, we present a multi-wavelength analysis of SN~2013dy, a
SN~Ia known to have the earliest detection \citep[$\sim2.4$\,hr after
the first light;][]{2013ApJ...778L..15Z}.  In addition to high-cadence
optical and NIR light curves, over 30 low-resolution optical spectra,
and one of the largest set of high-resolution optical spectra of a
SN~Ia, our dataset includes 10 UV spectra observed with the {\it
  Hubble Space Telescope} ({\it HST}).  We use this high-quality
dataset that spans epochs from \about 0.1 to 500\,days
after the explosion to constrain the properties of the explosion and
progenitor system for SN~2013dy.

This manuscript is organized as follows. In \S~\ref{sec:data}, we
present the photometric and spectroscopic observations of SN~2013dy.
In \S~\ref{sec:phot-analysis} and \ref{sec:spec-analysis}, we
present the photometric and spectroscopic properties of SN~2013dy,
respectively.  We compare models to our data in
\S~\ref{sec:model} and summarize our findings in
\S~\ref{sec:conclusion}.  Throughout this paper, we assume
$\mathrm{H_{0}} = 70$\,km\,s$^{-1}$\,Mpc$^{-1}$ and a flat universe
with $\omatter = 0.3$. 

\section{Observations}
\label{sec:data}
SN~2013dy was discovered in NGC~7250 \citep[$D =
13.7$\,Mpc;][]{2009AJ....138..323T} on 2013 July 10.46 (UT dates
are used throughout this paper) and classified
as a young SN~Ia \citep{2013CBET.3588....1C}.  Radio observations
taken \about 1 week after maximum brightness resulted in
nondetections \citep{2013ATel.5619....1P}.

\citet{2013ApJ...778L..15Z} obtained early-time photometry of
SN~2013dy starting \about 0.1\,d after explosion.  These data
constrained the progenitor star to be a compact object ($R_{0}
\leq 0.25\,{\rm R}_{\odot}$).  Their earliest spectrum had very strong
\ion{C}{2} lines, tracing the unburned material in the outer ejecta.
We combine the photometric and spectroscopic data from
\citet{2013ApJ...778L..15Z} with our own data presented below.
All of the data will be made available in the WISeREP archive
\citep{2012PASP..124..668Y}.

\subsection{Photometry}
\label{sec:phot}
\begin{figure}
	\centering
		\includegraphics*[scale=0.85]{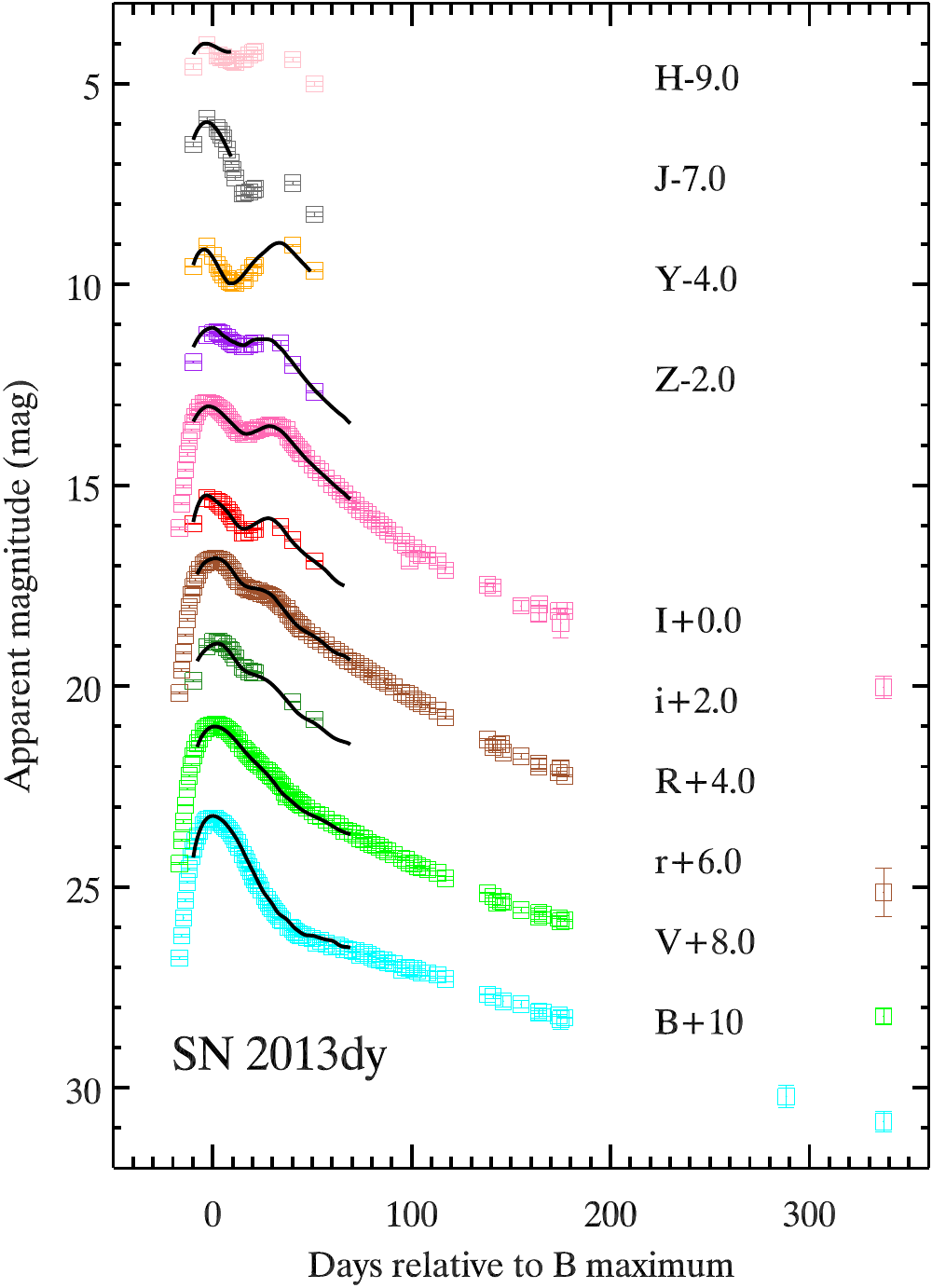}
                \caption{The KAIT \bvri\ and RATIR \rizyjh\ light curves
                of SN~2013dy. The solid lines represent the best fit (from 
                {\tt SNooPy}) to the observed light curves.}
        \label{sn2013dy_lc_all}
\end{figure}

We obtained broadband \bvri\ photometry of SN~2013dy with the 0.76-m
Katzman Automatic Imaging Telescope
\citep[KAIT;][]{2001ASPC..246..121F}.  The multi-band images were
observed with the KAIT4 filter set from $-16$\,d to $+337$\,d
relative to $B$-band maximum ($\rm MJD=56501.105$).
The data were reduced using our image-reduction
pipeline \citep{2010ApJS..190..418G}.  The point-spread function (PSF)
photometry was performed using {\tt DOAPHOT}
\citep{1987PASP...99..191S}. The SN instrumental magnitudes were
calibrated to nearby Sloan Digital Sky Survey (SDSS) standard stars
and transformed into the Landolt system.  Details about the KAIT
photometry are provided by \citet{2013ApJ...778L..15Z}.  In
Table~\ref{kait-phot1} and \ref{kait-phot2} we list the KAIT
photometry used in this work.

We also obtained \rizyjh\ photometry of SN~2013dy with the
multi-channel Reionization And Transients InfraRed camera
\citep[RATIR;][]{2012SPIE.8446E..10B} mounted on the 1.5-m Johnson
telescope at the Mexican Observatorio Astrono\'mico Nacional on Sierra
San Pedro M\'artir in Baja California, M\'exico
\citep{2012SPIE.8444E..5LW}.  Typical observations include a series of
80-s exposures in the $ri$~bands and 60-s exposures in the
$ZYJH$~bands, with dithering between exposures.  RATIR's fixed IR
filters cover half of their respective detectors, automatically
providing off-target IR sky exposures while the target is observed in
the neighboring filter. Master IR sky frames are created from a
median stack of off-target images in each IR filter.  No off-target
sky frames were obtained on the optical CCDs, but the small galaxy
size and sufficient dithering allowed for a sky frame to be created
from a median stack of all the images in each filter.  Flat-field
frames consist of evening sky exposures. Given the lack of a cold
shutter in RATIR's design, IR dark frames are not available.
Laboratory testing, however, confirms that the dark current is
negligible in both IR detectors \citep{2012SPIE.8453E..1OF}.

The RATIR data were reduced, coadded, and analysed using standard CCD
and IR processing techniques in IDL and Python, utilizing the online
astrometry programs {\tt SExtractor} and {\tt
  SWarp}\footnote{SExtractor and SWarp can be accessed from
  http://www.astromatic.net/software.} \citep[see][and references
therein]{2014MNRAS.443.2887F}.  Calibration was performed using field
stars with reported fluxes in both 2MASS \citep{2006AJ....131.1163S}
and the SDSS Data Release 9 Catalogue \citep{2012ApJS..203...21A}.
Table~\ref{ratir-phot} lists the RATIR photometry. In
Figure~\ref{sn2013dy_lc_all}, we present both the KAIT and RATIR light
curves and the best-fit template light curves (see
\S~\ref{sec:lc-fit}).

\begin{table*}
\small
\caption{KAIT photometry of SN~2013dy.}
\scalebox{0.93}{
\begin{tabular}{lcccc}
\hline\hline
MJD & $B$  & $V$  & $R$  & $I$\\
    & (mag) & (mag) & (mag) & (mag)\\
\hline
56484.42 & $16.76 \pm 0.04$ & $16.41 \pm 0.03$ & $16.17 \pm 0.03$ & $16.07 \pm 0.04$ \\
56485.46 & $16.21 \pm 0.03$ & $15.84 \pm 0.03$ & $15.60 \pm 0.03$ & $15.46 \pm 0.03$ \\
56486.45 & $15.77 \pm 0.07$ & $15.37 \pm 0.05$ & $15.17 \pm 0.03$ & $15.03 \pm 0.03$ \\
56487.38 & $15.33 \pm 0.03$ & $14.96 \pm 0.02$ & $14.74 \pm 0.02$ & $14.62 \pm 0.02$ \\
56488.40 & $14.88 \pm 0.03$ & $14.56 \pm 0.03$ & $14.34 \pm 0.03$ & $14.23 \pm 0.04$ \\
56489.37 & $14.55 \pm 0.03$ & $14.24 \pm 0.03$ & $14.02 \pm 0.03$ & $13.91 \pm 0.04$ \\
56490.43 & $14.24 \pm 0.02$ & $13.95 \pm 0.02$ & $13.72 \pm 0.02$ & $13.63 \pm 0.02$ \\
56491.40 & $14.04 \pm 0.03$ & $13.74 \pm 0.02$ & $13.53 \pm 0.03$ & $13.44 \pm 0.04$ \\
56492.39 & $13.86 \pm 0.03$ & $13.57 \pm 0.03$ & $13.35 \pm 0.03$ & $13.29 \pm 0.03$ \\
56493.40 & $13.73 \pm 0.06$ & $13.42 \pm 0.04$ & $13.22 \pm 0.03$ & $13.17 \pm 0.04$ \\
56494.41 & $13.61 \pm 0.03$ & $13.31 \pm 0.03$ & $13.11 \pm 0.03$ & $13.07 \pm 0.03$ \\
56496.32 & $13.46 \pm 0.05$ & $13.15 \pm 0.03$ & $12.98 \pm 0.02$ & $12.97 \pm 0.02$ \\
56497.41 & $13.37 \pm 0.03$ & $13.06 \pm 0.02$ & $12.93 \pm 0.02$ & $12.95 \pm 0.03$ \\
56498.38 & $13.33 \pm 0.03$ & $13.03 \pm 0.02$ & $12.90 \pm 0.02$ & $12.96 \pm 0.02$ \\
56499.38 & $13.33 \pm 0.07$ & $12.99 \pm 0.05$ & $12.87 \pm 0.06$ & $12.96 \pm 0.06$ \\
56500.45 & $13.28 \pm 0.03$ & $12.96 \pm 0.03$ & $12.83 \pm 0.03$ & $12.97 \pm 0.03$ \\
56501.39 & $13.30 \pm 0.03$ & $12.96 \pm 0.03$ & $12.84 \pm 0.03$ & $13.00 \pm 0.04$ \\
56502.35 & $13.30 \pm 0.03$ & $12.96 \pm 0.03$ & $12.83 \pm 0.03$ & $13.02 \pm 0.03$ \\
56503.30 & $13.33 \pm 0.03$ & $12.96 \pm 0.03$ & $12.84 \pm 0.03$ & $13.06 \pm 0.03$ \\
56504.35 & $13.37 \pm 0.03$ & $12.98 \pm 0.03$ & $12.84 \pm 0.03$ & $13.09 \pm 0.03$ \\
56505.34 & $13.40 \pm 0.03$ & $13.00 \pm 0.02$ & $12.88 \pm 0.02$ & $13.14 \pm 0.02$ \\
56506.35 & $13.44 \pm 0.02$ & $13.03 \pm 0.02$ & $12.90 \pm 0.02$ & $13.19 \pm 0.02$ \\
56507.33 & $13.50 \pm 0.04$ & $13.07 \pm 0.03$ & $12.93 \pm 0.04$ & $13.24 \pm 0.04$ \\
56508.33 & $13.54 \pm 0.02$ & $13.09 \pm 0.02$ & $12.99 \pm 0.02$ & $13.30 \pm 0.03$ \\
56509.34 & $13.59 \pm 0.03$ & $13.13 \pm 0.03$ & $13.05 \pm 0.03$ & $13.38 \pm 0.03$ \\
56510.34 & $13.66 \pm 0.03$ & $13.18 \pm 0.02$ & $13.11 \pm 0.03$ & $13.43 \pm 0.03$ \\
56511.31 & $13.73 \pm 0.03$ & $13.21 \pm 0.02$ & $13.15 \pm 0.02$ & $13.48 \pm 0.02$ \\
56512.31 & $13.82 \pm 0.03$ & $13.30 \pm 0.02$ & $13.26 \pm 0.03$ & $13.59 \pm 0.03$ \\
56513.32 & $13.90 \pm 0.03$ & $13.34 \pm 0.03$ & $13.33 \pm 0.02$ & $13.64 \pm 0.03$ \\
56514.32 & $14.01 \pm 0.03$ & $13.42 \pm 0.03$ & $13.39 \pm 0.03$ & $13.69 \pm 0.04$ \\
56516.33 & $14.19 \pm 0.03$ & $13.54 \pm 0.03$ & $13.50 \pm 0.03$ & $13.73 \pm 0.03$ \\
56517.32 & $14.30 \pm 0.04$ & $13.60 \pm 0.03$ & $13.54 \pm 0.03$ & $13.72 \pm 0.03$ \\
56518.36 & $14.39 \pm 0.03$ & $13.65 \pm 0.02$ & $13.56 \pm 0.02$ & $13.72 \pm 0.02$ \\
56519.31 & $14.51 \pm 0.03$ & $13.72 \pm 0.03$ & $13.60 \pm 0.02$ & $13.72 \pm 0.02$ \\
56520.34 & $14.60 \pm 0.03$ & $13.75 \pm 0.03$ & $13.59 \pm 0.02$ & $13.66 \pm 0.03$ \\
56521.41 & $14.74 \pm 0.15$ & $13.80 \pm 0.11$ & $13.60 \pm 0.06$ & $13.62 \pm 0.08$ \\
56522.33 & $14.80 \pm 0.03$ & $13.86 \pm 0.02$ & $13.63 \pm 0.02$ & $13.64 \pm 0.02$ \\
56523.25 & $14.90 \pm 0.03$ & $13.89 \pm 0.02$ & $13.64 \pm 0.02$ & $13.62 \pm 0.02$ \\
56524.31 & $14.96 \pm 0.07$ & $13.95 \pm 0.05$ & $13.64 \pm 0.05$ & $13.57 \pm 0.04$ \\
56525.29 & $15.07 \pm 0.04$ & $13.99 \pm 0.03$ & $13.68 \pm 0.03$ & $13.58 \pm 0.03$ \\
56527.35 & $15.24 \pm 0.04$ & $14.09 \pm 0.03$ & $13.71 \pm 0.03$ & $13.56 \pm 0.03$ \\
56528.36 & $15.32 \pm 0.03$ & $14.12 \pm 0.02$ & $13.72 \pm 0.03$ & $13.54 \pm 0.03$ \\
56529.33 & $15.40 \pm 0.03$ & $14.15 \pm 0.03$ & $13.74 \pm 0.03$ & $13.51 \pm 0.03$ \\
56530.33 & $15.48 \pm 0.03$ & $14.22 \pm 0.02$ & $13.78 \pm 0.02$ & $13.52 \pm 0.03$ \\
56531.32 & $15.54 \pm 0.03$ & $14.25 \pm 0.03$ & $13.79 \pm 0.03$ & $13.50 \pm 0.03$ \\
56532.35 & $15.62 \pm 0.03$ & $14.31 \pm 0.02$ & $13.84 \pm 0.02$ & $13.51 \pm 0.02$ \\
56533.30 & $15.67 \pm 0.03$ & $14.35 \pm 0.03$ & $13.87 \pm 0.03$ & $13.51 \pm 0.03$ \\
56534.31 & $15.73 \pm 0.03$ & $14.41 \pm 0.02$ & $13.93 \pm 0.02$ & $13.55 \pm 0.02$ \\
56535.28 & $15.80 \pm 0.03$ & $14.46 \pm 0.02$ & $13.98 \pm 0.02$ & $13.57 \pm 0.03$ \\
56536.26 & $15.85 \pm 0.03$ & $14.50 \pm 0.02$ & $14.02 \pm 0.02$ & $13.58 \pm 0.03$ \\
56537.29 & \nodata & $14.63 \pm 0.10$ & \nodata & \nodata \\
56538.28 & \nodata & $14.76 \pm 0.12$ & $14.09 \pm 0.15$ & $13.63 \pm 0.09$ \\
56539.28 & $16.01 \pm 0.03$ & $14.69 \pm 0.03$ & $14.21 \pm 0.03$ & $13.79 \pm 0.03$ \\
56540.29 & $16.00 \pm 0.03$ & $14.70 \pm 0.02$ & $14.25 \pm 0.02$ & $13.85 \pm 0.02$ \\
56541.26 & $16.04 \pm 0.04$ & $14.77 \pm 0.03$ & $14.34 \pm 0.03$ & $13.93 \pm 0.03$ \\
56542.28 & $16.10 \pm 0.03$ & $14.83 \pm 0.02$ & $14.38 \pm 0.02$ & $13.99 \pm 0.02$ \\
56543.26 & $16.13 \pm 0.03$ & $14.86 \pm 0.02$ & $14.43 \pm 0.02$ & $14.04 \pm 0.03$ \\
56544.24 & $16.17 \pm 0.03$ & $14.90 \pm 0.03$ & $14.47 \pm 0.03$ & $14.10 \pm 0.03$ \\
56545.26 & $16.20 \pm 0.04$ & $14.94 \pm 0.02$ & $14.54 \pm 0.02$ & $14.18 \pm 0.02$ \\
56546.26 & $16.19 \pm 0.04$ & $14.97 \pm 0.02$ & $14.56 \pm 0.03$ & $14.22 \pm 0.03$ \\
56548.24 & $16.24 \pm 0.03$ & $15.03 \pm 0.02$ & $14.64 \pm 0.02$ & $14.33 \pm 0.02$ \\
56551.26 & $16.28 \pm 0.04$ & $15.13 \pm 0.03$ & $14.75 \pm 0.03$ & $14.48 \pm 0.03$ \\
56553.24 & $16.38 \pm 0.06$ & $15.21 \pm 0.03$ & $14.84 \pm 0.02$ & $14.60 \pm 0.03$ \\
56555.27 & $16.34 \pm 0.05$ & $15.23 \pm 0.03$ & $14.88 \pm 0.03$ & $14.67 \pm 0.03$ \\
56558.23 & $16.41 \pm 0.04$ & $15.32 \pm 0.02$ & $14.98 \pm 0.02$ & $14.82 \pm 0.03$ \\
56561.23 & $16.49 \pm 0.04$ & $15.41 \pm 0.02$ & $15.09 \pm 0.02$ & $14.98 \pm 0.02$ \\
56563.22 & $16.45 \pm 0.03$ & $15.44 \pm 0.02$ & $15.14 \pm 0.03$ & $15.04 \pm 0.03$ \\
56565.22 & $16.49 \pm 0.03$ & $15.50 \pm 0.03$ & $15.21 \pm 0.03$ & $15.13 \pm 0.03$ \\
56567.23 & $16.54 \pm 0.06$ & $15.60 \pm 0.03$ & $15.28 \pm 0.03$ & $15.23 \pm 0.04$ \\
\hline
\label{kait-phot1}
\end{tabular}
}
\end{table*}

\begin{table*}
\small
\caption{KAIT photometry of SN~2013dy (continued).}
\scalebox{0.93}{
\begin{tabular}{lcccc}
\hline\hline
MJD & $B$  & $V$  & $R$  & $I$\\
    & (mag)& (mag)& (mag)& (mag)\\ 
\hline
56569.19 & $16.56 \pm 0.05$ & $15.61 \pm 0.03$ & $15.37 \pm 0.02$ & $15.35 \pm 0.03$ \\
56571.20 & $16.58 \pm 0.04$ & $15.67 \pm 0.03$ & $15.39 \pm 0.03$ & $15.40 \pm 0.03$ \\
56573.20 & $16.59 \pm 0.05$ & $15.70 \pm 0.03$ & $15.45 \pm 0.03$ & $15.49 \pm 0.04$ \\
56575.21 & $16.68 \pm 0.05$ & $15.79 \pm 0.02$ & $15.52 \pm 0.02$ & $15.60 \pm 0.03$ \\
56577.20 & $16.64 \pm 0.04$ & $15.81 \pm 0.03$ & $15.56 \pm 0.03$ & $15.66 \pm 0.03$ \\
56579.18 & $16.69 \pm 0.05$ & $15.89 \pm 0.03$ & $15.63 \pm 0.03$ & $15.73 \pm 0.03$ \\
56581.18 & $16.73 \pm 0.06$ & $15.93 \pm 0.03$ & $15.71 \pm 0.03$ & $15.84 \pm 0.04$ \\
56583.19 & $16.79 \pm 0.07$ & $15.96 \pm 0.03$ & $15.75 \pm 0.03$ & $15.89 \pm 0.04$ \\
56585.19 & $16.81 \pm 0.06$ & $16.00 \pm 0.03$ & $15.82 \pm 0.03$ & $15.98 \pm 0.03$ \\
56587.18 & $16.82 \pm 0.04$ & $16.08 \pm 0.04$ & $15.88 \pm 0.03$ & $16.05 \pm 0.04$ \\
56589.18 & $16.88 \pm 0.04$ & $16.11 \pm 0.03$ & $15.93 \pm 0.03$ & $16.13 \pm 0.04$ \\
56592.16 & $16.91 \pm 0.05$ & $16.19 \pm 0.04$ & $16.01 \pm 0.03$ & $16.24 \pm 0.03$ \\
56596.19 & $17.07 \pm 0.05$ & $16.28 \pm 0.03$ & $16.15 \pm 0.03$ & $16.41 \pm 0.04$ \\
56598.14 & $17.00 \pm 0.04$ & $16.32 \pm 0.03$ & $16.19 \pm 0.03$ & $16.48 \pm 0.04$ \\
56600.14 & $17.02 \pm 0.16$ & $16.40 \pm 0.12$ & $16.24 \pm 0.21$ & $16.88 \pm 0.14$ \\
56602.13 & $17.04 \pm 0.04$ & $16.42 \pm 0.03$ & $16.32 \pm 0.03$ & $16.57 \pm 0.04$ \\
56604.13 & $17.08 \pm 0.05$ & $16.49 \pm 0.03$ & $16.38 \pm 0.03$ & $16.69 \pm 0.05$ \\
56606.13 & $17.14 \pm 0.05$ & $16.52 \pm 0.04$ & $16.43 \pm 0.04$ & $16.73 \pm 0.05$ \\
56609.12 & $17.14 \pm 0.07$ & $16.56 \pm 0.04$ & $16.49 \pm 0.05$ & $16.73 \pm 0.06$ \\
56614.14 & $17.19 \pm 0.09$ & $16.65 \pm 0.06$ & $16.61 \pm 0.05$ & $16.88 \pm 0.10$ \\
56618.09 & $17.29 \pm 0.08$ & $16.78 \pm 0.04$ & $16.78 \pm 0.04$ & $17.11 \pm 0.08$ \\
56639.13 & $17.67 \pm 0.02$ & $17.15 \pm 0.02$ & $17.33 \pm 0.02$ & $17.47 \pm 0.04$ \\
56642.13 & $17.73 \pm 0.06$ & $17.24 \pm 0.03$ & $17.53 \pm 0.08$ & $17.55 \pm 0.09$ \\
56644.10 & \nodata & $17.39 \pm 0.12$ & $17.46 \pm 0.13$ & \nodata \\
56646.07 & \nodata & $17.35 \pm 0.10$ & $17.45 \pm 0.09$ & \nodata \\
56647.10 & $17.84 \pm 0.14$ & $17.38 \pm 0.08$ & $17.65 \pm 0.11$ & \nodata \\
56656.11 & $17.92 \pm 0.10$ & $17.58 \pm 0.07$ & $17.75 \pm 0.07$ & $18.00 \pm 0.13$ \\
56665.11 & $18.13 \pm 0.13$ & $17.73 \pm 0.08$ & $17.92 \pm 0.13$ & $18.20 \pm 0.18$ \\
56665.12 & $18.10 \pm 0.03$ & $17.65 \pm 0.04$ & $18.02 \pm 0.05$ & $17.97 \pm 0.09$ \\
56667.15 & $18.13 \pm 0.10$ & $17.69 \pm 0.06$ & \nodata & \nodata \\
56675.12 & $18.20 \pm 0.02$ & $17.77 \pm 0.02$ & $18.15 \pm 0.03$ & $18.12 \pm 0.05$ \\
56676.10 & $18.29 \pm 0.25$ & $17.85 \pm 0.15$ & $18.04 \pm 0.18$ & $18.43 \pm 0.37$ \\
56678.11 & $18.26 \pm 0.02$ & $17.82 \pm 0.02$ & $18.24 \pm 0.03$ & $18.13 \pm 0.06$ \\
56789.46 & $20.21 \pm 0.28$ & \nodata & \nodata & \nodata\\
56838.39 & $20.84 \pm 0.24$ & $20.22 \pm 0.19$ & $21.13 \pm 0.59$ & $20.03 \pm 0.28$ \\
\hline
\label{kait-phot2}
\end{tabular}
}
\end{table*}

\begin{table*}
\caption{RATIR photometry of SN~2013dy.}
\scalebox{0.93}{
\begin{tabular}{lcccccc}
\hline\hline
MJD & $r$  & $i$  & $Z$  & $Y$ & $J$ & $H$\\
    & (mag)& (mag)& (mag)& (mag) & (mag) & (mag)\\ 
\hline
56491.30 & $ 13.86 \pm  0.02$ & $ 13.96 \pm  0.02$ & $ 13.93 \pm  0.02$ & $ 14.15 \pm  0.02$ & $ 14.43 \pm  0.05$ & $ 14.93 \pm  0.07$ \\
56498.28 & $ 13.01 \pm  0.02$ & $ 13.32 \pm  0.02$ & $ 13.26 \pm  0.02$ & $ 13.66 \pm  0.02$ & $ 13.79 \pm  0.05$ & $ 14.39 \pm  0.07$ \\
56501.27 & $ 12.89 \pm  0.02$ & $ 13.35 \pm  0.02$ & $ 13.22 \pm  0.02$ & $ 13.89 \pm  0.02$ & \nodata & \nodata \\
56503.30 & $ 12.90 \pm  0.02$ & $ 13.40 \pm  0.02$ & $ 13.18 \pm  0.02$ & $ 14.10 \pm  0.02$ & $ 14.02 \pm  0.05$ & $ 14.64 \pm  0.07$ \\
56504.33 & $ 12.88 \pm  0.02$ & $ 13.44 \pm  0.02$ & $ 13.20 \pm  0.02$ & $ 14.20 \pm  0.02$ & $ 14.09 \pm  0.05$ & $ 14.63 \pm  0.07$ \\
56505.45 & $ 12.93 \pm  0.02$ & $ 13.48 \pm  0.02$ & $ 13.24 \pm  0.02$ & $ 14.29 \pm  0.02$ & $ 14.19 \pm  0.05$ & $ 14.69 \pm  0.07$ \\
56506.45 & $ 12.95 \pm  0.02$ & $ 13.52 \pm  0.02$ & $ 13.29 \pm  0.02$ & $ 14.35 \pm  0.02$ & $ 14.28 \pm  0.05$ & $ 14.68 \pm  0.07$ \\
56508.32 & $ 13.03 \pm  0.02$ & $ 13.62 \pm  0.02$ & $ 13.34 \pm  0.02$ & $ 14.45 \pm  0.02$ & $ 14.55 \pm  0.05$ & $ 14.73 \pm  0.07$ \\
56509.44 & $ 13.09 \pm  0.02$ & $ 13.69 \pm  0.02$ & $ 13.40 \pm  0.02$ & $ 14.51 \pm  0.02$ & \nodata & \nodata \\
56510.43 & $ 13.12 \pm  0.02$ & $ 13.76 \pm  0.02$ & $ 13.44 \pm  0.02$ & $ 14.54 \pm  0.02$ & $ 14.89 \pm  0.05$ & $ 14.76 \pm  0.07$ \\
56511.35 & $ 13.21 \pm  0.02$ & $ 13.83 \pm  0.02$ & $ 13.46 \pm  0.02$ & $ 14.57 \pm  0.02$ & $ 15.05 \pm  0.05$ & $ 14.79 \pm  0.07$ \\
56512.39 & $ 13.30 \pm  0.02$ & $ 13.93 \pm  0.02$ & $ 13.49 \pm  0.02$ & $ 14.57 \pm  0.02$ & $ 15.26 \pm  0.05$ & $ 14.81 \pm  0.05$ \\
56516.26 & $ 13.53 \pm  0.02$ & $ 14.19 \pm  0.02$ & $ 13.55 \pm  0.02$ & $ 14.49 \pm  0.02$ & $ 15.66 \pm  0.05$ & $ 14.74 \pm  0.05$ \\
56517.24 & $ 13.56 \pm  0.02$ & $ 14.19 \pm  0.02$ & $ 13.54 \pm  0.02$ & $ 14.44 \pm  0.02$ & $ 15.64 \pm  0.05$ & $ 14.72 \pm  0.05$ \\
56519.47 & $ 13.63 \pm  0.02$ & $ 14.15 \pm  0.02$ & $ 13.49 \pm  0.02$ & $ 14.31 \pm  0.02$ & $ 15.62 \pm  0.05$ & $ 14.63 \pm  0.05$ \\
56521.42 & $ 13.66 \pm  0.02$ & $ 14.10 \pm  0.02$ & $ 13.47 \pm  0.02$ & $ 14.17 \pm  0.02$ & $ 15.55 \pm  0.05$ & $ 14.57 \pm  0.05$ \\
56522.30 & $ 13.67 \pm  0.02$ & $ 14.09 \pm  0.02$ & $ 13.47 \pm  0.02$ & $ 14.12 \pm  0.02$ & $ 15.53 \pm  0.05$ & $ 14.55 \pm  0.05$ \\
56535.30 & \nodata & $ 14.04 \pm  0.02$ & $ 13.46 \pm  0.05$ & \nodata & \nodata & \nodata \\
56541.30 & $ 14.39 \pm  0.02$ & $ 14.36 \pm  0.02$ & $ 14.00 \pm  0.02$ & $ 13.62 \pm  0.02$ & $ 15.39 \pm  0.05$ & $ 14.75 \pm  0.05$ \\
56545.30 & \nodata & \nodata & \nodata & \nodata & \nodata & \nodata \\
56552.30 & $ 14.82 \pm  0.02$ & $ 14.89 \pm  0.02$ & $ 14.68 \pm  0.02$ & $ 14.25 \pm  0.02$ & $ 16.17 \pm  0.05$ & $ 15.35 \pm  0.05$ \\
\hline
\label{ratir-phot}
\end{tabular}
}
\end{table*}

\subsection{Spectroscopy}
\label{sec:spec}
As part of our follow-up campaign, we obtained a series of
low-resolution near-UV through NIR spectra with {\it HST}, as well as
low- and high-resolution optical spectra from a variety of sources.

\begin{figure*}
	\centering
	\begin{tabular}{c}
		\includegraphics*[scale=0.75]{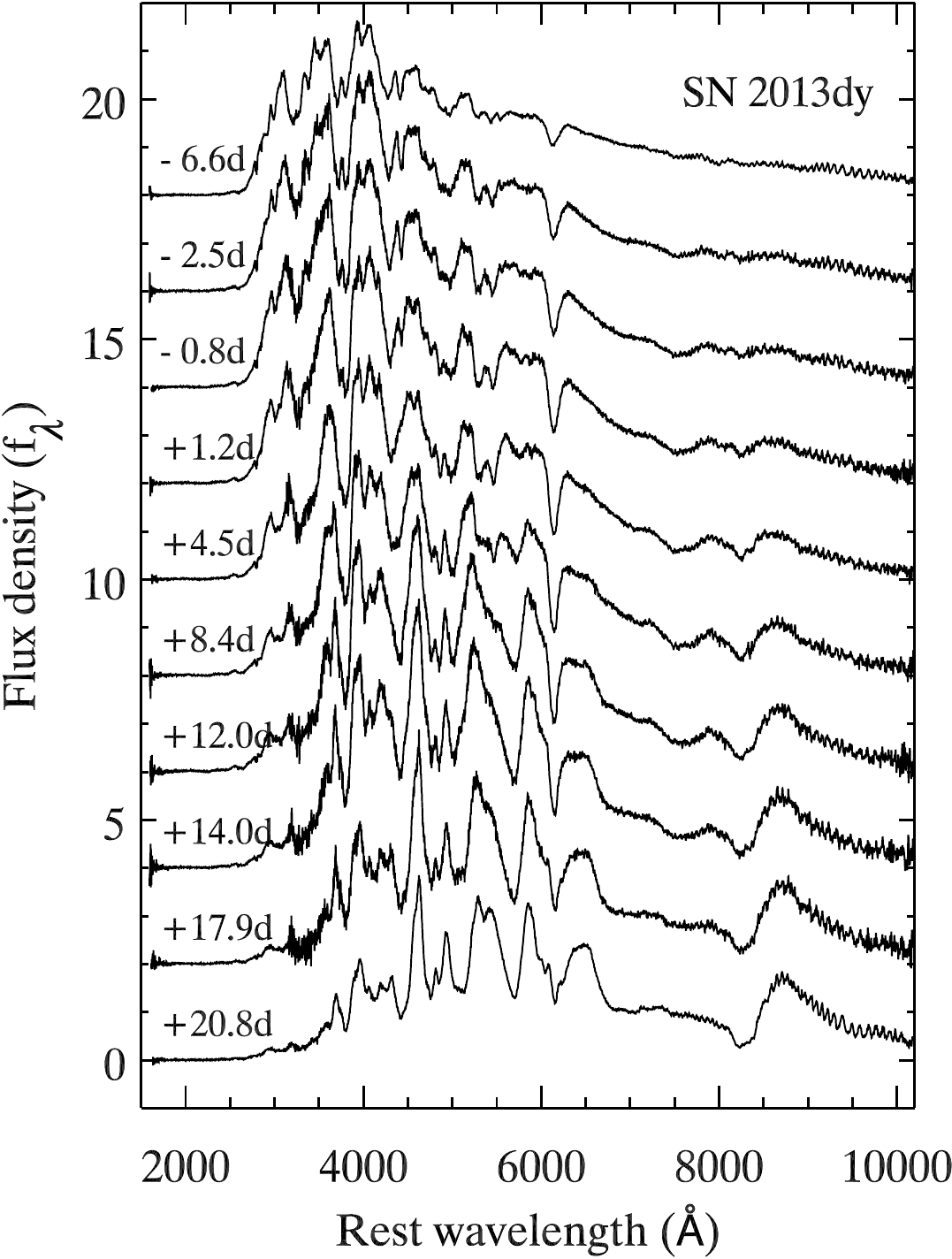} 
		\includegraphics*[scale=0.75]{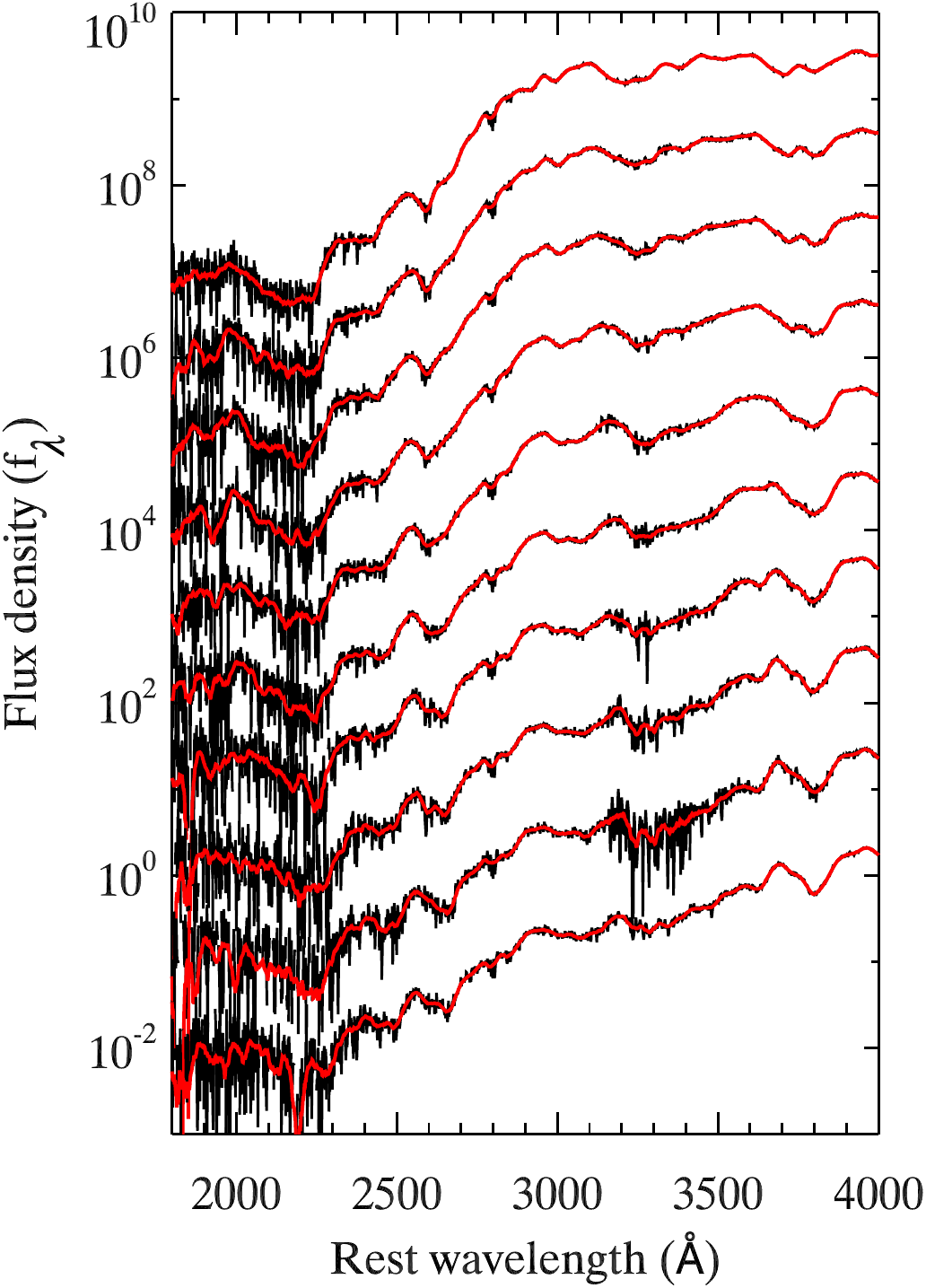}
	\end{tabular}
               \caption{Left: Spectra of SN~2013dy
               taken with {\it HST}/STIS, with arbitrary vertical
               offsets for clarity. Right: Same as left panel, 
               but a closer look at the UV region. The
               spectra smoothed with a Savitzky-Golay filter \citep{1964AnaCh..36.1627S}
               are shown in red curves.
               }
        \label{13dy-hst-spec}
\end{figure*}

We procured 10 epochs of near-UV through NIR spectroscopy with the
Space Telescope Imaging Spectrograph (STIS) on {\it HST}.  For each
epoch, we obtained data with three different setups to cover the
entire wavelength range: the NUV MAMA detector with the G230L grating,
the CCD detector with the G430L grating, and the CCD detector with the
G750L grating.  Combined, the three setups have a wavelength range of
1600--10,230\,\AA.  The {\it HST} data were reduced using the
standard Space Telescope Science Data Analysis System (STSDAS)
routines, including bias subtraction, flat-fielding, wavelength
calibration, and flux calibration \citep{2012ApJ...753L...5F}.  The
{\it HST} observations correspond to phases ranging from $-7$ to
$+21$\,d relative to $B$-band maximum brightness.  A log of our {\it
  HST} spectroscopic observations is presented in Table~\ref{hst-log}
and the spectra are shown in Figure~\ref{13dy-hst-spec}.

In addition to the {\it HST} spectroscopy, we obtained 32 optical
spectra of SN~2013dy with phases of $-16$ to $+480$\,d relative to
$B$-band maximum brightness with a variety of ground-based facilities.
The 8 earliest spectra in the dataset were already studied by
\citet{2013ApJ...778L..15Z}, but are included in our analysis for
completeness.  The new spectra (24 of the 32 total) were observed with
the Low Resolution Imaging Spectrometer
\citep[LRIS;][]{1995PASP..107..375O} mounted on the 10-m Keck-I
telescope, the DEep Imaging Multi-Object Spectrograph
\citep[DEIMOS;][]{2003SPIE.4841.1657F} on the 10-m Keck-II
telescope, the Kast Spectrograph
\citep{Kast_spectrograph} on the Lick Observatory 3-m Shane
telescope, and the Low-Resolution Spectrograph
\citep[LRS;][]{1998SPIE.3355..375H} on the 9.2-m Hobby-Eberly
Telescope (HET).

\begin{figure}
	\centering
	\begin{tabular}{c}
		\includegraphics*[scale=0.8]{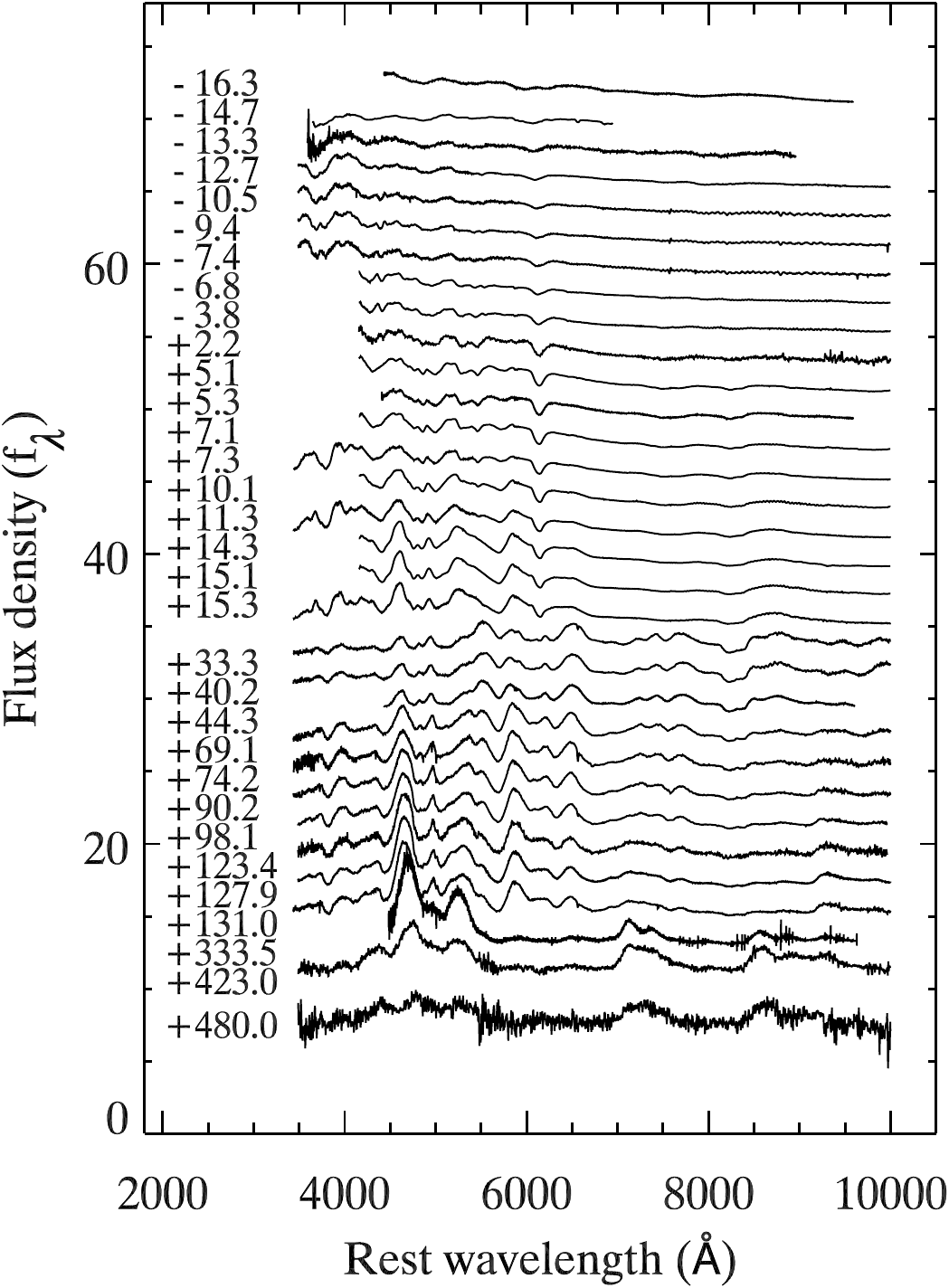}
	\end{tabular}
               \caption{The optical spectra of SN~2013dy.
               }
        \label{13dy-opt-spec}
\end{figure}

Standard CCD processing and spectrum extraction were accomplished with
IRAF.  The data were extracted using the optimal algorithm of
\citet{1986PASP...98..609H}.  Low-order polynomial fits to calibration-lamp
spectra were used to establish the wavelength scale, and small
adjustments derived from night-sky lines in the object frames were
applied.  We employed our own IDL routines to flux calibrate the data
and remove telluric lines using the well-exposed continua of the
spectrophotometric standard stars \citep{1988ApJ...324..411W, 2003PASP..115.1220F}.  
Details of our spectroscopic reduction techniques are described by
\citet{2012MNRAS.425.1789S}.
A log of spectroscopic observations is presented in
Table~\ref{opt-log}.  The complete spectral sequence can be found in
Figure~\ref{13dy-opt-spec}.

\begin{table}
\centering
\caption{Log of {\it HST} spectroscopic observations of SN~2013dy.}
\begin{tabular}{lcccc}
\hline\hline
Date		& Phase &	\multicolumn{3}{c}{Exposure time}\\
(MJD)		& (days) &   \multicolumn{3}{c}{(s)}\\
\hline
			&	      & G230L	&	G420L	&	G750L\\
\hline
56494.48	& $-6.6$  & 3618	&	336		&	336\\
56498.60	& $-2.5$  & 1382	&	64		&	64\\
56500.32	& $-0.8$  & 1382	&	64		&	64\\
56502.31	& $+1.2$  & 1382	&	64		&	64\\
56505.57	& $+4.5$  & 1382	&	64		&	64\\
56509.49	& $+8.4$  & 1382	&	64		&	64\\
56513.09	& $+12.0$ & 1382	&	64		&	64\\
56515.14	& $+14.0$ & 1382	&	64		&	64\\
56518.99	& $+17.9$ & 1382	&	64		&	64\\
56521.91	& $+20.8$ & 3618	&	336		&	336\\
\hline
\end{tabular}
\label{hst-log}
\end{table}

\begin{table}
\centering
\caption{Log of low-resolution optical spectroscopic observations of SN~2013dy.}
\begin{tabular}{lccc}
\hline\hline
Date & Phase & Instrument & Exp. time\\
(MJD)& (day) &            &  (s)\\
\hline
56497.28 & $-3.8$ & HET/LRS 		&  	380\\
56503.26 & $+2.2$ & HET/LRS 		&  	577\\
56506.24 & $+5.1$ & HET/LRS			&  	100\\
56506.44 & $+5.3$ & Keck/DEIMOS 	& 	60\\
56508.22 & $+7.1$ & HET/LRS 		&  	260\\
56508.41 & $+7.3$ & Lick/Kast 		& 	360\\
56511.21 & $+10.1$ & HET/LRS 		&  	400\\
56512.41 & $+11.3$ & Lick/Kast 		& 	360\\
56515.45 & $+14.3$ & HET/LRS 		&  	200\\
56516.20 & $+15.1$ & HET/LRS 		&  	200\\
56516.41 & $+15.3$ & Lick/Kast 		& 	360\\
56534.42 & $+33.3$ & Lick/Kast 		& 	180\\
56541.32 & $+40.2$ & Lick/Kast 		& 	180\\
56545.36 & $+44.3$ & Keck/DEIMOS 	& 	300\\
56570.23 & $+69.1$ & Lick/Kast 		& 	450\\
56575.26 & $+74.1$ & Lick/Kast 		& 	450\\
56591.33 & $+90.2$ & Lick/Kast 		& 	900\\
56599.24 & $+98.1$ & Lick/Kast 		& 	1800\\
56624.24 & $+123.1$ & Lick/Kast 	& 	1800\\
56629.22 & $+127.9$ & Keck/LRIS 	& 	350\\
56632.12 & $+131.0$ & Lick/Kast 	& 	1800\\
56834.58 & $+333.5$ & Keck/DEIMOS 	& 	2400\\
56924.34 & $+423.0$ & Keck/LRIS 	& 	1000\\
56981.32 & $+480.0$ & Keck/LRIS 	& 	1000\\
\hline
\end{tabular}
\label{opt-log}
\end{table}

We also observed SN~2013dy with various high-resolution spectrographs
on 15 different occasions spanning a phase range of $-10$ to
$+19$~days.  This corresponds to a mean (and median) cadence of 2~days,
with the largest gap between spectra being 5~days.  The
high-resolution spectra were obtained with the Tillinghast Reflector
Echelle Spectrograph (TRES) mounted on the 1.5-m Tillinghast
telescope, the High Efficiency and Resolution Mercator Echelle
Spectrograph \citep[HERMES;][]{2011A&A...526A..69R} on the
1.2-m Mercator Telescope, the High-Resolution Spectrograph
\citep[HRS;][]{1998SPIE.3355..387T} on the HET, the
Ultraviolet and Visual Echelle Spectrograph
\citep[UVES;][]{2000SPIE.4008..534D} on the Very Large
Telescope (VLT), the High Resolution Echelle Spectrometer
\citep[HRES;][]{1994SPIE.2198..362V} on the Keck-I telescope,
and the High Accuracy Radial velocity Planet Searcher
\citep[HARPS-N;][]{2012SPIE.8446E..1VC} on the Telescope
Nationale Galileo (TNG).  These data represent one of the largest single sets
of high-resolution spectra of a SN~Ia\footnote{There are several
  separate sets of high-resolution spectra of SN~2014J
   \citep{2014MNRAS.443.2887F, 2014ApJ...784L..12G,
    2014ApJ...792..106W, 2015ApJ...801..136G,2015arXiv150600938J}; 
    the largest single set of spectra has 33 epochs.  The largest single set of
  spectra of SN~2011fe has 12 epochs \citep{2013A&A...549A..62P}.}.
The spectra have resolutions ranging from $\sim$ 30,000 to 115,000. 
A log of
observations is given in Table~\ref{hres-log}.  The spectra were
reduced with standard procedures \citep[e.g.,][and references
therein]{2014MNRAS.443.1849S} including telluric correction.  For a
subset of the observations, we observed standard stars to
determine the telluric absorption in the SN~2013dy spectra; for the
remaining spectra, we produced synthetic telluric spectra
\citep{2015arXiv150107239S}.  We used B-splines to fit the continuum
and normalize the spectra.

\begin{table}
\centering
\caption{Log of high-resolution observations of SN~2013dy.}
\begin{tabular}{lcccc}
\hline\hline
Date & Phase & Instrument & $\rm T_{Exp.}$ & Resolution\\
(MJD)& (day) &            &  (s) &\\
\hline
56491.42 & $-9.7$ & Tillinghast/TRES 	&  	2831 &	30,000\\
56494.09 & $-7.0$ & Mercator/HERMES 	&  	3600 &	85,000\\
56494.26 & $-6.8$ & HET/HRS 			& 	1000 &	33,600\\
56495.13 & $-6.0$ & Mercator/HERMES 	&  	3600 &	85,000\\
56497.26 & $-3.8$ & HET/HRS 			& 	1000 &	33,600\\
56502.23 & $+1.1$ & VLT/UVES 			&  	600  &	40,000\\
56504.25 & $+3.1$ & HET/HRS 			& 	1000 &	33,600\\
56505.05 & $+3.9$ & Mercator/HERMES 	&  	2700 &	85,000\\
56506.61 & $+5.5$ & Keck/HIRES 			&  	300  &	48,000\\ 
56511.06 & $+10.0$& Mercator/HERMES 	& 	2700 &	85,000\\
56513.22 & $+12.1$& HET/HRS 			& 	1200 &	33,600\\
56515.97 & $+14.9$& TNG/HARPS	 		& 	1800 &	115,000\\
56516.20 & $+15.1$& HET/HRS 			& 	1200 &	33,600\\
56519.08 & $+18.0$& Mercator/HERMES 	& 	3600 &	85,000\\
56520.07 & $+19.0$& Mercator/HERMES 	& 	3600 &	85,000\\
\hline
\end{tabular}
\label{hres-log}
\end{table}

\section{Photometric Analysis}
\label{sec:phot-analysis}
In this section, we analyze the SN~2013dy photometry.  We fit the
light curves, deriving estimates of the host-galaxy reddening and
extinction, examine the color evolution, and construct a bolometric
light curve.  Using broadband photometry of the host galaxy, we
derive host parameters including stellar mass (\mstellar) and
star-formation rate (SFR).

\subsection{Light-Curve Fitting}
\label{sec:lc-fit}
\begin{table*}
\centering
\caption{Results of {\tt SNooPy} light-curve fitting for SN~2013dy$^a$.}
\begin{tabular}{ccccc}
\hline\hline
$T_{\rm max}$ & $B_{\rm max}$ & $\Delta m_{15}$ & DM & $E(B-V)_{\rm host}$\\
(MJD)      &   (mag)   &     (mag)    &  (mag) &  (mag)\\
\hline
56501.105  & $13.229\pm 0.010$ & $0.886\pm 0.006$ & $31.488 \pm 0.010$ & $0.206\pm 0.005$\\
\hline
\end{tabular}
\label{snoopy-res}
\\$^a${The statistical uncertainties of the measurements are generated by Monte-Carlo simulations.}
\end{table*}

We fit the SN~2013dy \bvrriizyjh\ light curves simultaneously with
{\tt SNooPy} \citep{2011AJ....141...19B}. {\tt SNooPy} is a
Python-based light-curve fitter, extending the method of
\citet{2006ApJ...647..501P}. The decline-rate parameter, $\Delta
m_{15}$, \citep[similar to $\Delta m_{15}(B)$ defined
by][corresponding to the $B$-band decline 15 days after maximum
brightness]{1993ApJ...413L.105P} is used to parameterize the SN
light-curve shape in {\tt SNooPy}.  Given that our dataset were observed
with different photometric systems, we provide {\tt SNooPy} with corresponding 
filter functions and  photometric zero-points to perform K- and S-corrections.
We adopted the default ``EBV model''
in {\tt SNooPy} to fit the light curves, which is described in detail
below.  The fitting results in measurements of $\Delta m_{15}$, time
of $B$-band maximum brightness, distance modulus (DM) 
and the host-galaxy reddening $E(B-V)_{\rm host}$.

For the EBV model, a $B$-band peak absolute magnitude (\absm) as well as the
colors are assumed based on the value of $\Delta m_{15}$.
Six different calibrations are provided in the model. 
The parameters of these calibrations were derived by \citet{2010AJ....139..120F} 
to minimize the Phillips relation \citep[\absm\ versus $\Delta m_{15}$;][]{1993ApJ...413L.105P} 
using different subsets of the training SN sample, with different $R_{V}$ determined
(or fixed) for each calibration.

  We tested all six available calibrations and found that
  calibration \#2 \citep[see Table~9 in][]{2010AJ....139..120F} with the EBV model, 
  which corresponds to $R_{V} = 3.1$, produces the best fitting of the SN~2013dy light curves (in
  terms of $\chi^2$, with $\chi^2_{\nu} = 2.93$).  We adopted this
  setup when fitting the light curves.  The results of the light-curve
  fitting are shown in Table~\ref{snoopy-res}.  The light-curve
  fitting with different reddening models gives consistent $\Delta
  m_{15}$, but lower $E(B-V)_{\rm host}$ for those with low $R_{V}$
  ($E(B-V)_{\rm host}=0.16$ and 0.18\,mag for $R_{V}=1.46$ and 1.01,
  respectively).  This leads to a lower extinction measurement
  compared to that derived from a $R_{V}=3.1$ reddening model.
  However, the low-$R_{V}$ models produce much poorer fits to the
  light curves, especially in the NIR, than $R_{V}=3.1$.  The setup
  with $R_{V}=3.1$ is still strongly preferred for SN~2013dy even if
  we only fit the \bvri\ light curves (where the data were better
  sampled).  Previous studies have found that a low $R_{V}$ value is
  preferred for some highly-reddened SNe~Ia
  \citep[e.g.,][]{2006AJ....131.1639K,2008ApJ...675..626W}. On the other 
  hand, we find that the dust reddening of SN~2013dy is much more likely to
  have a ``high'' value of $R_{V}$ ($R_{V} \approx 3.1$) than it is a
  ``low'' value ($R_{V} < 2$).  For the rest of the study, we
  therefore use the EBV--$R_{V} = 3.1$ model for our
  primary results, but examine how the conclusions change with a lower
  value of $R_{V}$ when appropriate.  

{\tt SNooPy} measures a decline-rate parameter $\Delta m_{15}=0.89$.
This template-derived $\Delta m_{15}$ is similar to the conventional
$B$-band decline rate $\Delta m_{15}(B)$, but with some systematic differences.
Using the relation provided by \citet{2011AJ....141...19B}, we calculate
a $B$-band decline rate $\Delta m_{15}(B)=0.92$\,mag. This indicates
SN~2013dy has a relatively slower decline rate, compared to the normal SN~Ia
like SN~2011fe \citep[$\Delta m_{15}(B) \approx1.1$\,mag;][]{2013NewA...20...30M},
but is similar to another slow decliner SN~1991T \citep[$\Delta m_{15}(B) \approx
0.94$\,mag;][]{1996AJ....112.2438H}. 

The host-galaxy reddening, $E(B-V)_{\rm host} = 0.206 \pm 0.005$\,mag
measured by {\tt SNooPy}, is similar to that measured by
\citet[$E(B-V)_{\rm host}=0.15$\,mag]{2013ApJ...778L..15Z} using the
equivalent-width (EW) of the \nai\,D absorption as a proxy for
reddening.  We adopt our value and the \citet*[][CCM]{1989ApJ...345..245C} 
reddening law throughout the analyses in this work. 
Our best-fit reddening corresponds to a host-galaxy visual
extinction of $A_{V, {\rm host}} = 0.64 \pm 0.02$\,mag. For the
Milky Way reddening, we adopt $E(B-V)_{\rm MW} = 0.14$\,mag
\citep{1998ApJ...500..525S, 2011ApJ...737..103S} and $R_{V} = 3.1$.

{\tt SNooPy} also determines a distance modulus $\rm DM=31.49$\,mag.
This gives a luminosity distance of 19.84\,Mpc for SN~2013dy, which is much greater
than the Tully-Fisher distance to NGC~7250 of $13.7 \pm 3.0$\,Mpc \citep{2009AJ....138..323T}.
However, the distance in {\tt SNooPy} is derived by fitting the SN light curves, so any parameters that
affect the light-curve properties (e.g., metallicity) could also affect the distance.
Given that the purpose of this work is to investigate how these parameters could affect
the SN properties, it would be important that we use a SN-independent distance in the analysis. 
Accordingly, we adopt the Tully-Fisher distance to NGC~7250 throughout this work.

\subsection{Color Curves}
\label{sec:color-curves}
\begin{figure}
	\centering
	\begin{tabular}{c}
		\includegraphics*[scale=0.8]{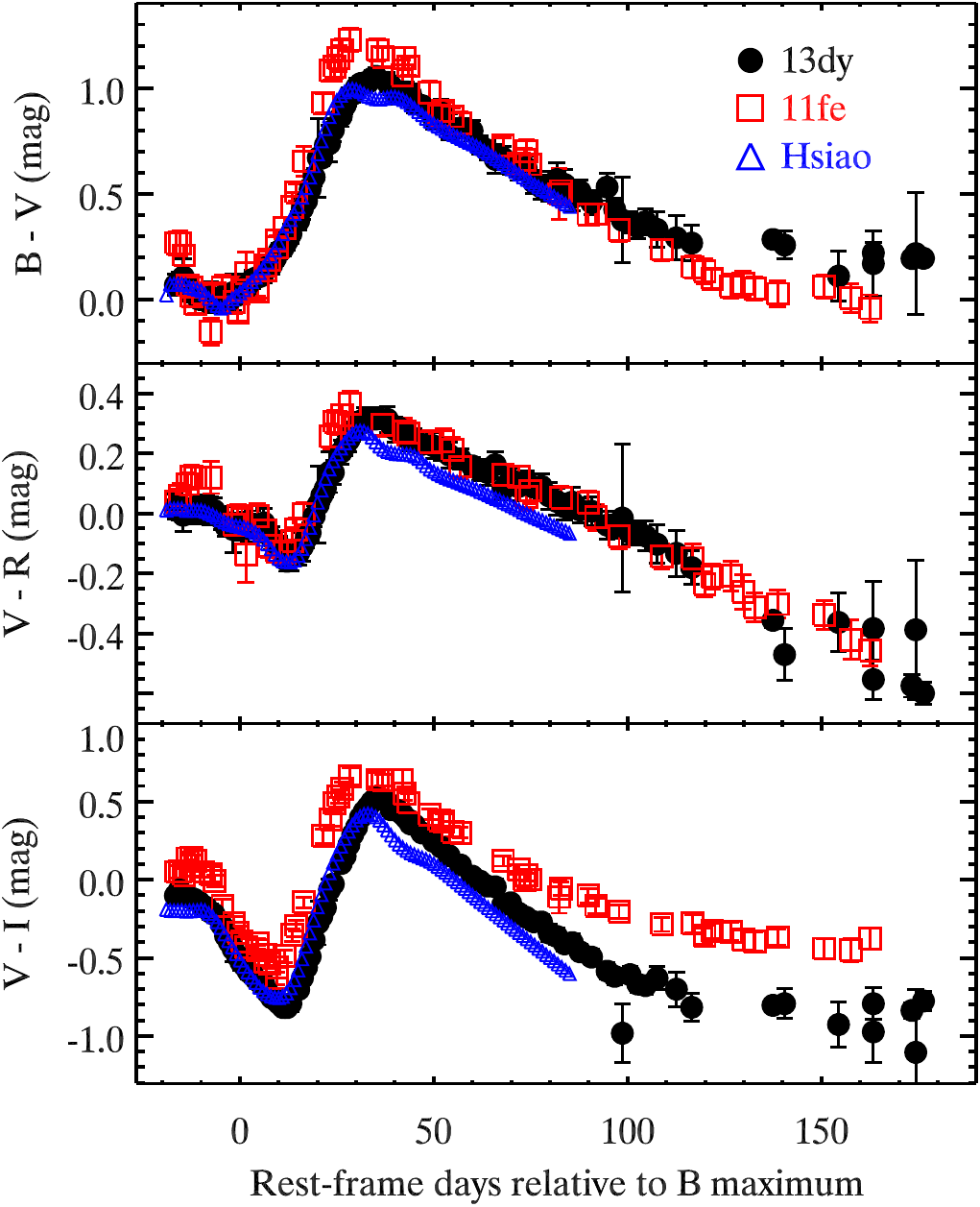}
	\end{tabular}
               \caption{The color curves of SN~2013dy.
               From top to bottom: The $B-V$, $V-R$, and $V-I$ color as
               as the function of phase. 
               The filled circles are the data in this work. 
               The color curves from SN~2011fe (red open squares) and the 
               \citet{2007ApJ...663.1187H} templates (blue open triangles)
               are compared.
               }
        \label{opt-col-curve}
\end{figure}

\begin{figure}
	\centering
	\begin{tabular}{c}
		\includegraphics*[scale=0.8]{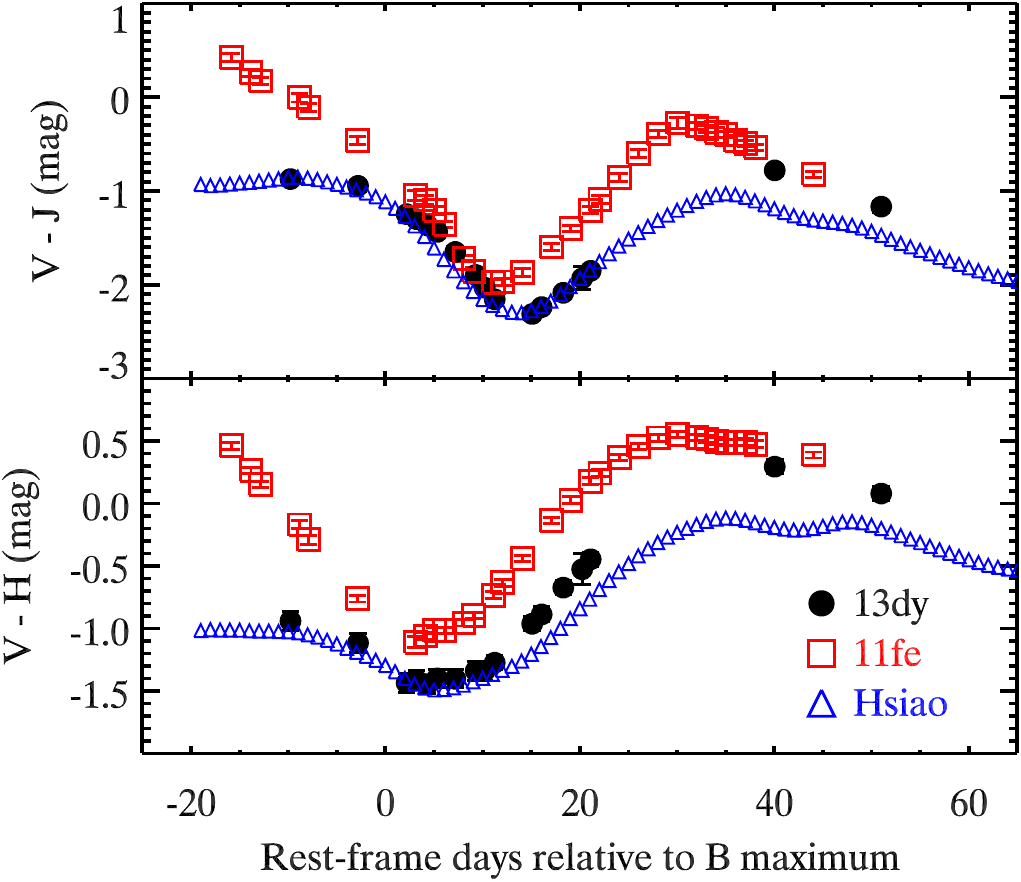}
	\end{tabular}
               \caption{The same as Figure~\ref{opt-col-curve},
               but using $V-J$ and $V-H$ color instead.              
               }
        \label{nir-col-curve}
\end{figure}

The dereddened color curves of SN~2013dy are shown in
Figures~\ref{opt-col-curve} and \ref{nir-col-curve}.  For comparison,
we also display the color curves of SN~2011fe and the
\citet{2007ApJ...663.1187H} templates.  The
\citet{2007ApJ...663.1187H} color curves are shifted to match
SN~2013dy at maximum brightness. For SN~2011fe, we adopted the $BVRI$ and $JHK$
photometry studied by \citet{2012JAVSO..40..872R} and
\citet{2012ApJ...754...19M}, respectively. Here the dereddened color curves
of SN~2011fe are directly compared without applying any shifts.

We find that after dereddening by our adopted reddening parameters,
the SN~2013dy and SN~2011fe $B-V$ and $V-R$ color curves are generally
consistent.  However, SN~2013dy is slightly bluer (weighted-mean $\Delta (B - V) =
-0.07\pm0.01$\,mag and $\Delta (V - R) = -0.06\pm0.01$\,mag) than SN~2011fe 
at early phases ($t \leq 30$\,d). The color curves of SN~2013dy 
also peak at later times ($\Delta t \approx 3$~days for $B-V$) and have shallower 
slopes at later phases ($t \gtrsim 30$\,d) than SN~2011fe.  
\citet{2014ApJ...789...32B} showed that the
time of $B-V$ maximum is strongly correlated with $\Delta m_{15}(B)$,
with faster-declining SNe~Ia having earlier $B-V$ maxima and steeper
late-time $B-V$ slopes than SNe~Ia with slower decline rates. Given
that SN~2013dy is a slower decliner ($\Delta m_{15}(B) = 0.9$\,mag) than
SN~2011fe ($\Delta m_{15}(B) \approx1.1$\,mag), our results are consistent with
the predicted trends.

The color curves of SNe~2011fe and 2013dy have larger differences at
longer wavelengths.  We find that SN~2013dy is bluer than SN~2011fe in
$V-I$ at all epochs.  The same trends are also found for the $V-J$ and
$V-H$ colors, with even greater differences.  Finally, relative to
SN~2011fe, SN~2013dy has much less premaximum $V-J$ and $V-H$
color evolution.

The color curves of the \citet{2007ApJ...663.1187H} templates are
generally similar to SN~2013dy at early times. However, they fail to
reproduce the trends at $t \gtrsim 30$\,d, where the templates are
bluer than SN~2013dy.

\subsection{Bolometric Light Curve}
\label{sec:bolometric}
\begin{figure*}
	\centering
	\begin{tabular}{c}
		\includegraphics*[scale=0.75]{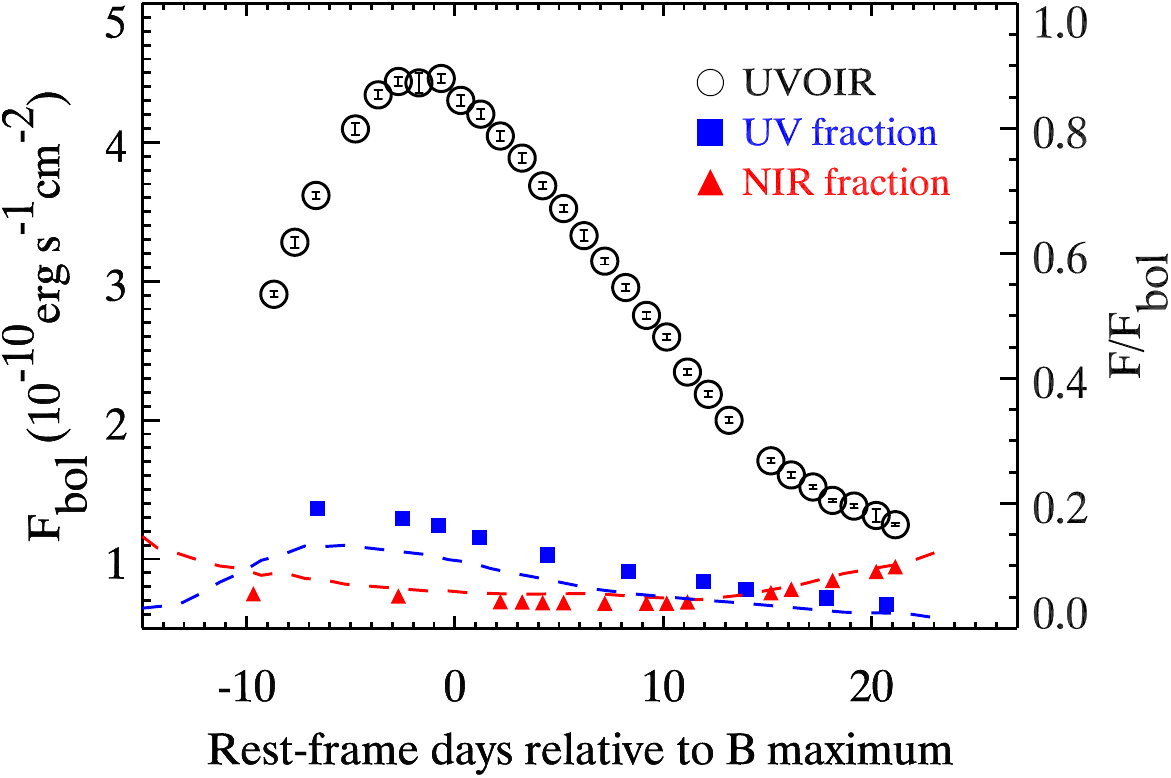}
		\includegraphics*[scale=0.75]{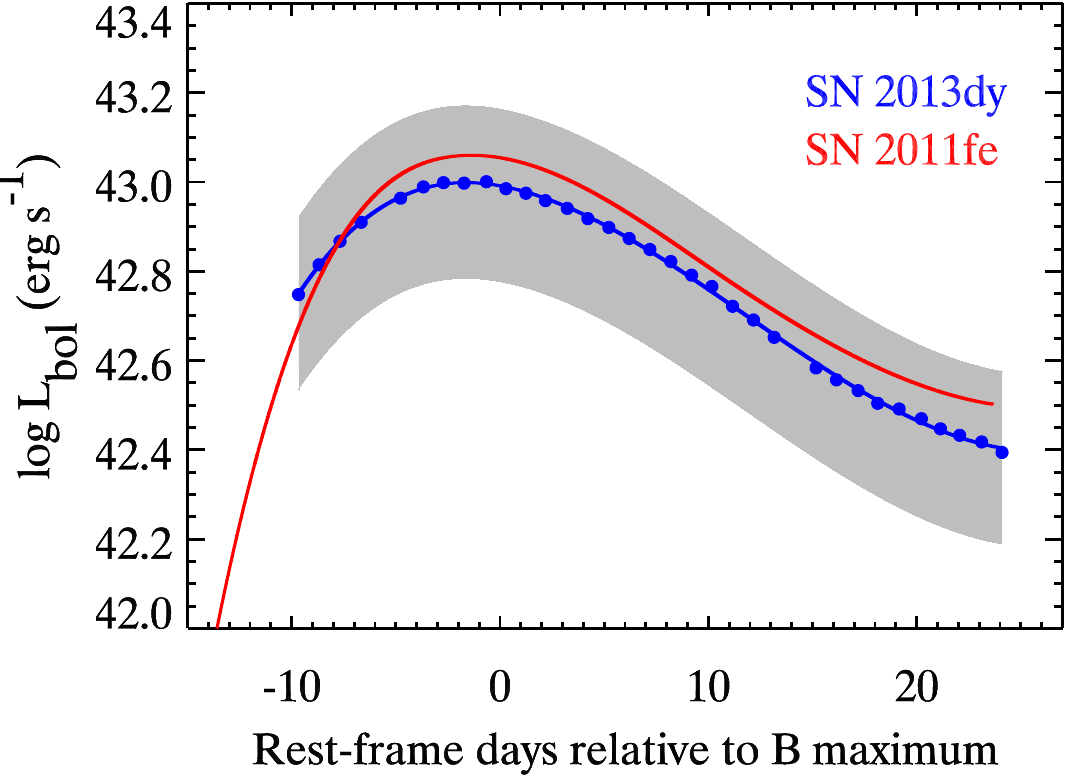}
	\end{tabular}
               \caption{Left: The UVOIR bolometric flux of SN~2013dy (open circles).
               The UV flux fractions determined by {\it HST} UV spectroscopy
               are represented by blue filled squares. The NIR flux
               fractions determined by RATIR NIR spectroscopy are
               represented by red filled triangles. The blue and red dashed
               line represents the UV and NIR flux fractions of SN~2011fe
               \citep{2013A&A...554A..27P}, respectively. Right: The same as left panel,
               but using bolometric luminosity instead. The UVOIR bolometric luminosity of SN~2013dy
               is shown in blue filled circles. Here the luminosity distance
               determined using Tully-Fisher relation \citep[$13.7\pm3.0$\,Mpc;][]{2009AJ....138..323T}
               is adopted. The blue curve represents the B-spline fit to the data. 
               The grey area shows the 1$\sigma$ range of the bolometric luminosities
               considering the uncertainties from the distance.
               The bolometric luminosity of SN~2011fe from \citet{2013A&A...554A..27P} 
               is overplotted (red curve).
               }
        \label{bolometric-lc}
\end{figure*}

With our exquisite optical/NIR photometry and multi-epoch UV
spectroscopy, we are able to construct a pseudo-bolometric light curve
which covers 1600--18,000\,\AA, which covers essentially the entire
SN emission.  Since the amount of light beyond these wavelengths is
likely a very small fraction of all emission, we will refer to the
pseudo-bolometric light curve as a ``bolometric'' light curve, but
recognize that there may be small differences with the true bolometric
light curve.  For all measurements, we correct for Milky Way and
host-galaxy reddening, using the parameters derived in
\S~\ref{sec:lc-fit}.

The KAIT \bvri\ optical light curves, which cover \about 3400--9700\,\AA, 
cover phases from $-17$\,d to $+340$\,d.  Because of the
high cadence of the KAIT observations, we do not use the RATIR $ri$
light curves to construct the bolometric light curve.  Integrating the
total flux through all filters, we construct a \bvri\ optical light
curve.  While the bulk of the SN emission is in the optical, a
significant amount of light is emitted in the UV and NIR.

Using the 10 epochs of {\it HST} spectroscopy (covering phases of
$-6.6$ to $+20.8$\,d), we can construct a UV light curve covering 
$\sim1600$--3400\,\AA.  The UV flux is calculated by integrating the SN
spectrum from 1600\,\AA\ to the effective wavelength of the $B$ band at
that particular epoch.  For the 10 epochs with UV spectra, we
determine the total UV$+$optical flux and the corresponding fraction
of light emitted in the UV.  Linearly interpolating the fraction of UV
light, we estimate the fraction of light emitted in the UV for all
epochs of the \bvri\ optical light curve between the first and last
epochs of UV spectroscopy.

A NIR light curve was constructed using the \yjh\ RATIR light curves.  These
data cover $\sim9700$--18,000\,\AA.  We did not include the RATIR $Z$-band
photometry since a large portion of the filter overlaps with the KAIT
$I$ band.  The NIR light curve was generated using a procedure similar
to that of the optical light curve.  Since our dataset does not cover
the flux in the $K$ band, we assume a linear decline from the $H$-band
effective wavelength to zero flux at 18,000\,\AA.  Similar to what was
done for the UV light curve, the fraction of the NIR flux in the
optical$+$NIR light curve is calculated.  Again, linearly
interpolating the NIR fraction, the corresponding NIR flux at each
individual epoch of the optical light curve (for those in the phase
range covered by the NIR photometry) is determined.

The final UVOIR bolometric light curve was generated by combining
the optical light curve with the fraction of light determined to be
emitted at UV and NIR wavelengths.  The result is shown in the left
panel of Figure~\ref{bolometric-lc}.  Note that the bolometric light
curve in this work is only valid from $t \approx -10$\,d to $+22$\,d,
the phases where both {\it HST} UV spectroscopy and RATIR NIR
photometry are available.

Figure~\ref{bolometric-lc} also presents the fraction of the total
light emitted in the UV and NIR.  We find that the UV contribution
peaks at the earliest epochs, with 19\% of the bolometric light being
emitted at UV wavelengths at $t \approx -10$~d.  The fraction of light
emitted in the UV declines steeply, becoming only 4\% of the total at
$+21$\,d.

The NIR fraction shows a different evolution from the UV fraction, being
6\% of all light at $-10$\,d, dropping to a minimum of 4\% at
$+10$\,d, and then increasing to 10\% at $+21$\,d.  The NIR
contribution overtakes that of the UV at  \about $+15$~d.

For comparison, we also show the fractions calculated for SN~2011fe
\citep{2013A&A...554A..27P}.  The overall trend of the UV fraction
decreasing and the NIR fraction increasing with time also occurs for
SN~2011fe.  However, SN~2013dy emits more light in the UV and less
light in the NIR than SN~2011fe at all phases.  At its peak, SN~2011fe
emitted 13\% of its light in the UV (compared to 19\% for SN~2013dy).
At $+20$~d, the UV fraction for both SNe is similar (2\% and 4\% for
SNe~2011fe and 2013dy, respectively).  This trend of more UV emission
for SN~2013dy with diminishing difference with time is also seen
directly in the UV spectra (see \S~\ref{sec:uv-spec}).  For
SN~2011fe, the NIR contributes 9\% of the total light at $-10$~d,
decreasing some to be 5\% at $+10$\,d and then increasing to 10\% at
$+20$\,d, again making SNe~2011fe and 2013dy be similar at the later
phases.

The bolometric luminosity of SN~2013dy is shown in the right panel of
Figure~\ref{bolometric-lc}.  Here, we use the Tully-Fisher distance to
NGC~7250 \citep[$13.7 \pm 3.0$\,Mpc;][]{2009AJ....138..323T}.  Fitting
a B-spline to the bolometric light curve, we derive a peak bolometric
luminosity of $L_{\rm bol, peak} = 10.0^{+4.8}_{-3.8} \times
10^{42}$~erg~s$^{-1}$, where the uncertainty in the peak bolometric 
luminosity is dominated by the uncertainty in the distance to SN~2013dy.
If the distance derived in {\tt SNooPy} (19.8\,Mpc; see \S~\ref{sec:lc-fit}) is adopted,
we derive a peak bolometric luminosity of 
$L_{\rm bol, peak} = 2.09\times10^{43}$~erg~s$^{-1}$, which is $\sim2$ times larger than
that derived from the Tully-Fisher distance.

We calculate $\Delta m_{15}$ from the bolometric light curve by fitting the
light curve with a B-spline. We find that SN~2013dy has bolometric
$\Delta m_{15}=0.98$\,mag, 
which is larger than its $B$-band decline rate $\Delta m_{15}(B)$.
Using the same method, we determine $\Delta m_{15}=0.99$\,mag for SN~2011fe.
This suggests that SN~2013dy and SN~2011fe have nearly identical bolometric
decline rates. \citet{2013ApJ...778L..15Z} determined a rise time (here the rise 
time is defined as the time elapsed from first-light to $B$-band maximum brightness) 
of $\sim 17.7$ days for SN~2013dy. This value is consistent to that of SN~2011fe
\citep[17.6 days;][]{2015MNRAS.446.3895F}, and close to the average of normal
SNe Ia \citep[17.4 days;][]{2010ApJ...712..350H}. We find there is hint that 
SN~2013dy is likely to have longer rise time than SN~2011fe
based on the bolometric light-curves from $t\sim-10$\,d to maximum brightness.
However, the bolometric rise time of SN~2013dy is difficult to constrain directly
from the bolometric light-curve due to the lack of UV and NIR observations in early times. 
Using the time of first light derived by \citet{2013ApJ...778L..15Z}, we
find a bolometric rise time for SN~2013dy of $\sim 16.47$ days. Assuming the Tully-Fisher
distance of NGC~7250 is correct, this will result in a \Nifs\ mass estimate of 
$\sim 0.44\,\rm M_{\sun}$ for SN~2013dy, using Eq.~(6) of \citet{2005A&A...431..423S}.

\subsection{Host Galaxy}
\label{sec:host}
\begin{table}
\centering
\caption{The host-galaxy photometry used in this work.}
\begin{tabular}{ccccc}
\hline\hline
SDSS $u$ & SDSS $g$ & SDSS $r$ & SDSS $i$ & SDSS $z$\\
(mag)    &   (mag)  &   (mag)  &   (mag)  &   (mag) \\  
\hline
13.225(4) & 12.654(2) & 12.352(2) & 12.635(2) & 12.390(4)\\
\hline
\end{tabular}
\label{host-phot}
\end{table}

\begin{table}
\centering
\caption{Summary of host-galaxy parameters derived by {\tt Z-PEG}.}
\begin{tabular}{llccc}
\hline\hline
 	&	& Lower limit & Best & Upper limit\\
\hline
$\log\rm \mstellar$	&($\rm M_{\sun}$)		&	8.48	&	8.67	&	8.84\\
$\log\rm SFR$		&($\rm M_{\sun}\,yr^{-1}$)	&	$-0.47$	&	$-0.26$	&	$-0.06$\\
$\log\rm sSFR$	&($\rm yr^{-1}$)			&	$-8.94$	&	$-8.92$	&	$-8.90$\\
\hline
\end{tabular}
\label{zpeg-log}
\end{table}

The host galaxy of SN~2013dy, NGC~7250, is a blue, late-type (likely
between Sc and Im) galaxy.  Using SDSS \ugriz\ photometry
\citep{2014ApJS..211...17A} of NGC~7250 (listed in Table~\ref{host-phot}), 
we derive physical parameters for the host galaxy.  
The details of the procedure are described by
\citet{2014MNRAS.438.1391P}.  Here we briefly summarize the process.

The host-galaxy stellar mass (\mstellar) and SFR
are determined using the photometric redshift code {\tt Z-PEG}
\citep{2002A&A...386..446L}.  {\tt Z-PEG} fits the observed color of
the galaxies with galaxy SEDs from 9
different spectral types (SB, Im, Sd, Sc, Sbc, Sb, Sa, S0, and E).  We
correct for Milky Way extinction, and allow for a further foreground
dust screen varying from $E(B-V) = 0$ to 0.2\,mag in steps of
0.02\,mag.  A \citet{1955ApJ...121..161S} initial-mass function (IMF)
is assumed.

The best-fit \mstellar\ and SFR determined by {\tt Z-PEG} are $\log\rm
(M/M_{\sun}) = 8.67$ and 0.55\,$\rm M_{\sun}\,yr^{-1}$, respectively.
This indicates that NGC~7250 has a relatively low \mstellar\
\citep[e.g., the Milky Way has a $\log\rm (M/M_{\sun}) =
10.79$;][]{2006MNRAS.372.1149F}.  Based on the mass--metallicity
relation derived by \citet{2004ApJ...613..898T}, we inferred the
gas-phase metallicity of NGC~7250 to be $12+\log\rm(O/H) = 8.49$. This
value is well below the solar value
\citep[8.69;][]{2001ApJ...556L..63A}, but only slightly larger than
the metallicity of the Large Magellanic Cloud
\citep[8.40;][]{1999IAUS..190..266G}.  Spectroscopic observations are
necessary to precisely constrain the metallicity of the host galaxy.
Using the \mstellar\ derived above, we can then determine the
SFR relative to the underlying galaxy stellar mass, which
is called specific SFR (sSFR), the SFR per unit \mstellar\
\citep{1997ApJ...489..559G}. Expressing sSFR in units of
yr$^{-1}$, this gives a $\log\rm sSFR =
-8.92$ for NGC~7250, which is a particularly large value (the
Milky Way has a $\log\rm sSFR = -10.18$;
\citealt{2006Natur.439...45D}) and provides evidence that there are
strong star-forming regions in the host galaxy. A summary of the {\tt
  Z-PEG} results is given in Table~\ref{zpeg-log}.

\section{Spectroscopic analysis}
\label{sec:spec-analysis}
In this section, we analyze the spectroscopic data.  We measure
pseudo-equivalent widths (pEWs) and velocities of key spectral features
in the UV, optical, and NIR, and study their temporal evolution.
The variations of narrow absorption features are also examined.  We
investigate the spectral properties of SN~2013dy by comparing them 
with those of a large sample of SNe~Ia.

\subsection{Photospheric Optical Spectroscopy}
\label{sec:opt-nir-spec}
\begin{figure}
	\centering
	\begin{tabular}{c}
		\includegraphics*[scale=0.82]{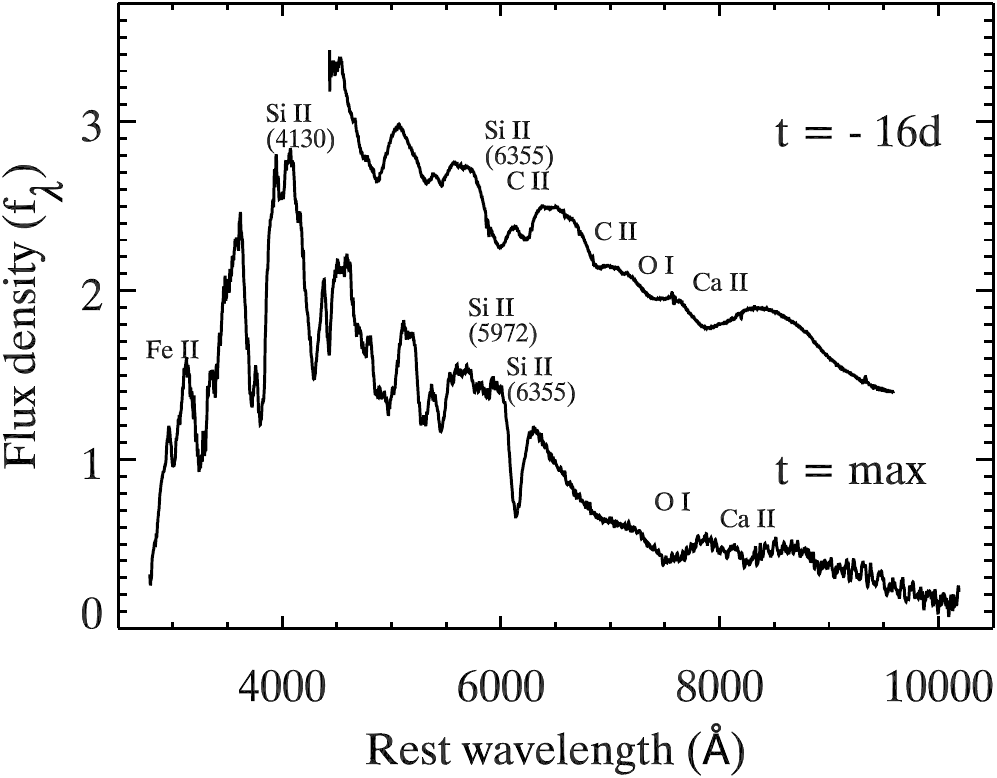}
	\end{tabular}
               \caption{Spectra at $t=-16$\,d and $+0$\,d demonstrating
               the key spectral features measured in this work.
               }
        \label{spec_example}
\end{figure}

\begin{figure}
	\centering
	\begin{tabular}{c}
		\includegraphics*[scale=0.75]{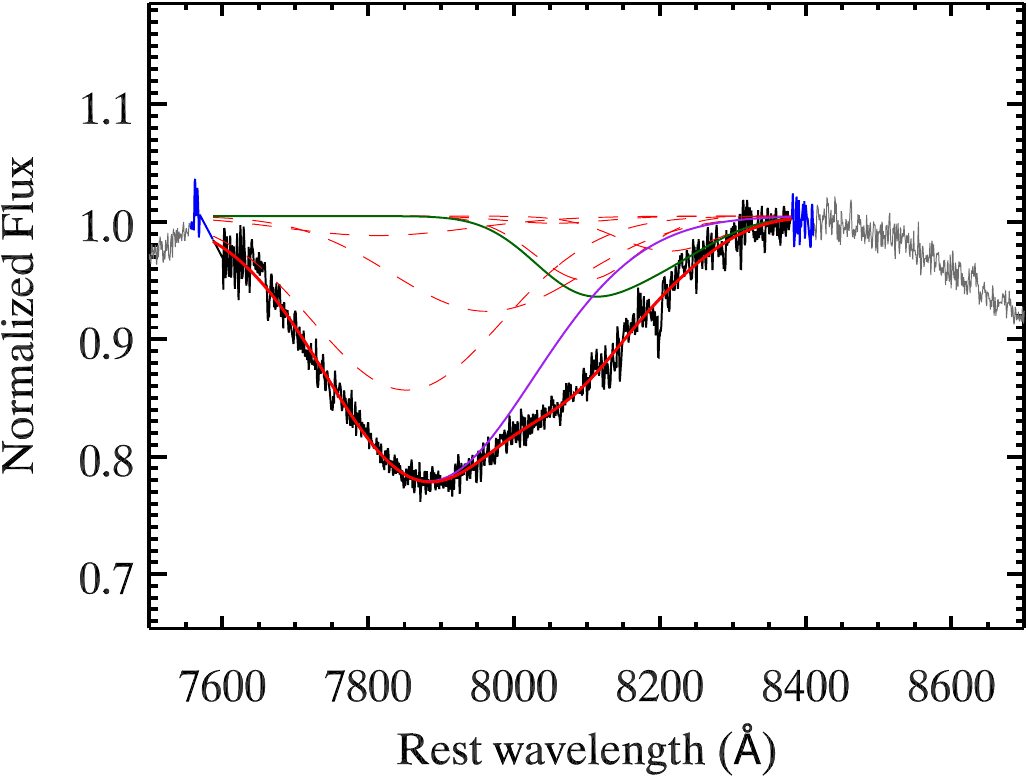}\\
		\includegraphics*[scale=0.75]{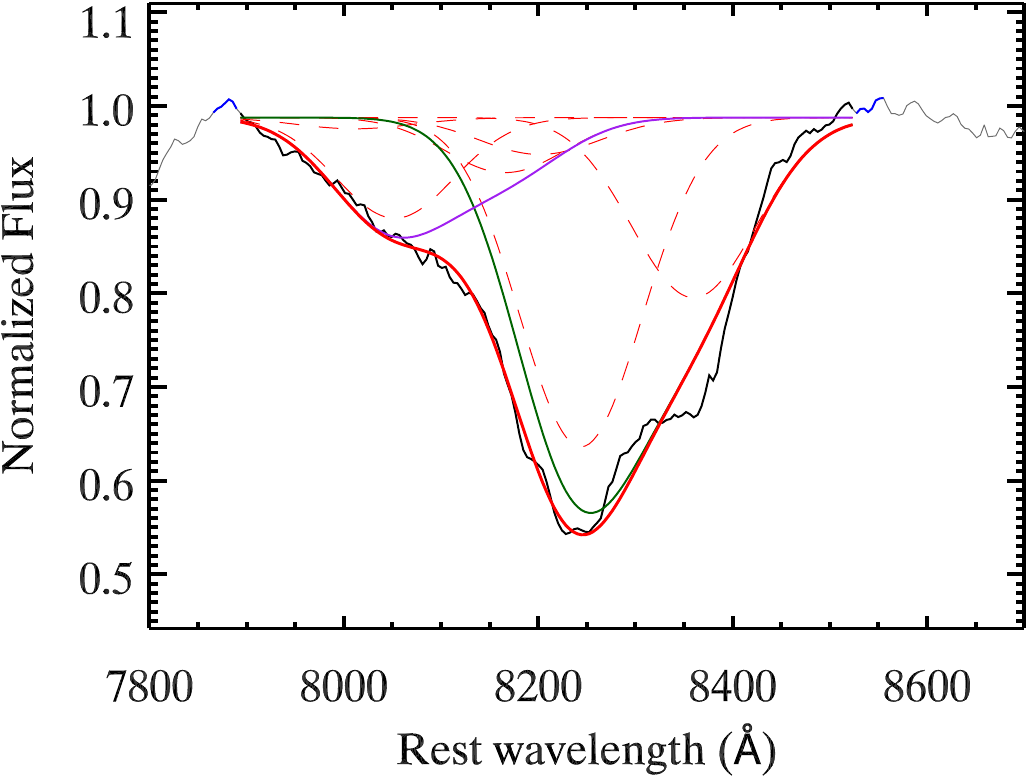}\\
		\includegraphics*[scale=0.75]{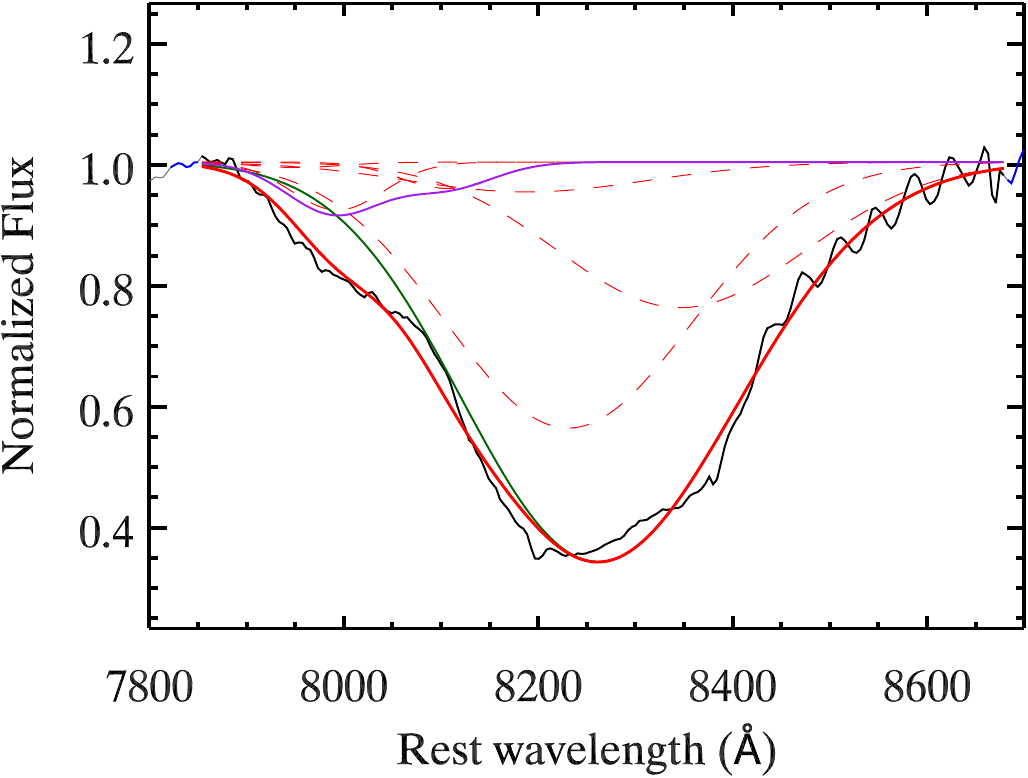}
	\end{tabular}
               \caption{From top to bottom: The \Caii\ NIR
               absorption at $-16$, $+5$ and $+15$ days,
               respectively. The black curve shows the
               range we fit the line profile. The blue continuum
               regions near the left and right side of the
               line profile are the regions we select to fit
               the pseudo-continuum. The red solid line is
               the best fit to the line profile, which is the 
               superposition of all the line components used for
               the fitting (represented in dashed lines).
               The green curve represents the superposition of 
               all the components in \Caii\ NIR photospheric-velocity
               feature (PVF). The purple curve represents the
               superposition of all the components in the
               \Caii\ NIR high-velocity feature (HVF).
               }
        \label{cair_evolution2}
\end{figure}

\begin{figure}
	\centering
	\begin{tabular}{c}
		\includegraphics*[scale=0.75]{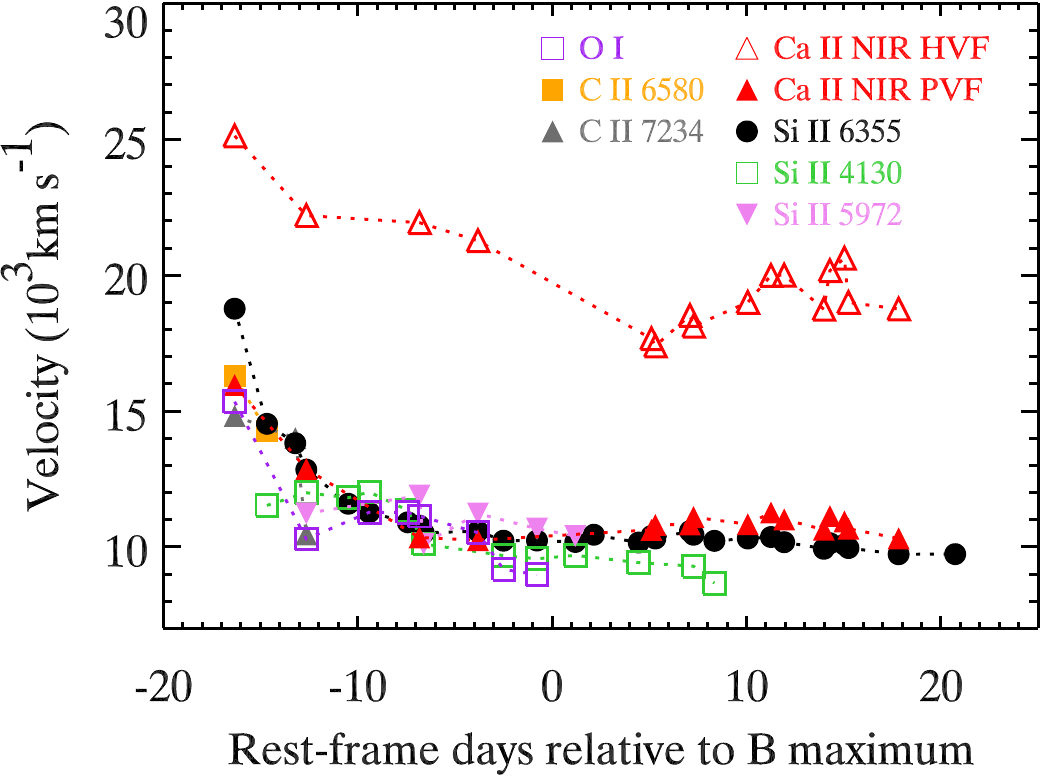}\\
		\includegraphics*[scale=0.75]{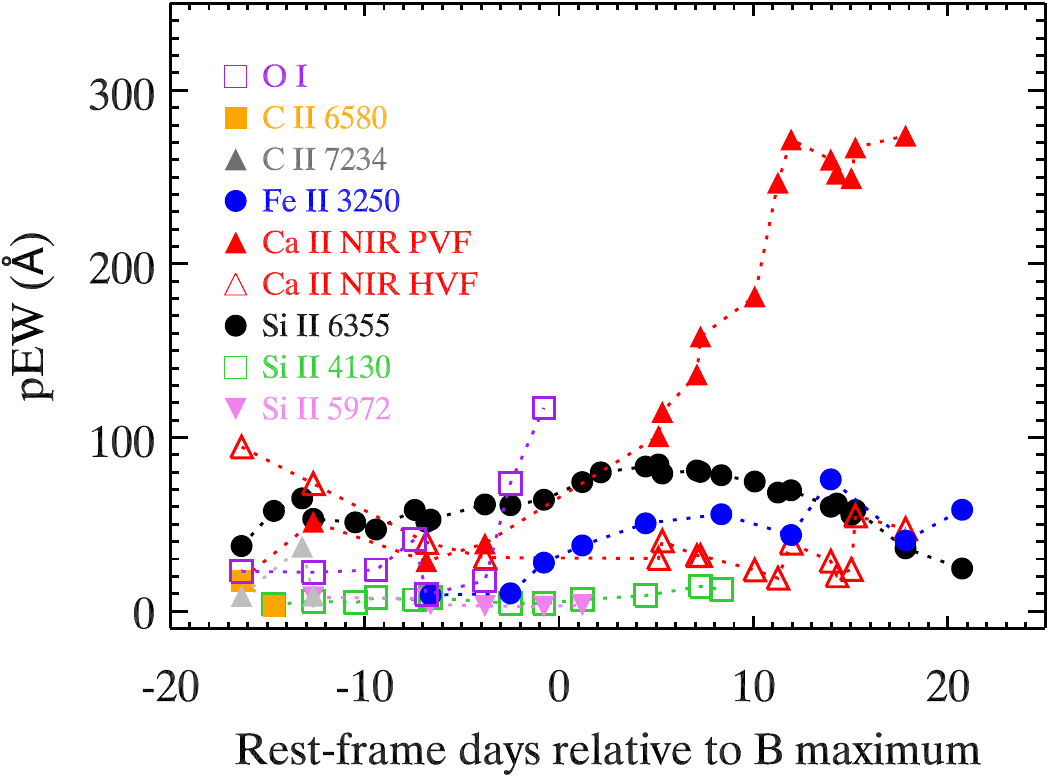}
	\end{tabular}
               \caption{Top: The SN ejecta velocity as a function of phase for
               various spectral features. Bottom: Same as top panel,
               but using pEWs instead. 
               }
        \label{spec_evolution}
\end{figure}

The well-observed spectral sequence of SN~2013dy makes it an ideal
target to study individual spectral features and their temporal
evolution.  Using the method outlined by \citet{2014MNRAS.444.3258M}
and \citet{2015MNRAS.446..354P}, we measured several spectral
features.  The key features of interest are the pEWs
and velocities of the \oi\ triplet, \cii\ and \ciitmp,
\Siii, \Siiitmp, \Siiitmpsec, and the \Caii\ NIR
triplet. Here we measured \Caii\ NIR instead of \Caii\ H\&K since it
provides us cleaner measurements of \Caii\ velocity and pEW without
contamination from other features \citep{2013MNRAS.435..273F,
2014MNRAS.437..338C, 2014MNRAS.444.3258M}.  In Figure~\ref{spec_example} 
we show examples of our photospheric spectra with the relevant spectral features marked.
Here we are able to decompose the high-velocity feature (HVF) and photospheric velocity
feature (PVF) in the \Caii\ NIR triplet (see Figure~\ref{cair_evolution2} for examples). 

Figure~\ref{spec_evolution} shows the velocities and pEWs of the
spectral features as a function of phase. \citet{2013ApJ...778L..15Z}
discovered the presence of both \cii\ and \ciitmp\ absorption in the
earliest spectra, corresponding to 16 days before $B$ maximum
brightness, which shows evidence of unburned material in the
outer layers of the ejecta.  In this work we confirm the detection of
$\rm{C}\,\textsc{ii}$ and measure physical values similar to that of
\citet{2013ApJ...778L..15Z}.  The \cii\ and \ciitmp\ features were
found with velocities of 16,300\,km\,$\rm s^{-1}$ and
14,800\,km\,$\rm s^{-1}$ at $t = -16$\,d, respectively.  However, both
features quickly faded away after $t = -13$\,d (\about 3 days after
explosion).  The \oi\ triplet can also be clearly identified in the
spectral sequence, with an initial velocity of 15,300\,km\,$\rm
s^{-1}$ at $t = -16$\,d and decreasing to 8,900\,km\,$\rm
s^{-1}$ at maximum light.

The velocity of \Siii\ is \about 18,800\,km\,$\rm s^{-1}$ at $-16$\,d,
consistent with the value reported by \citet{2013ApJ...778L..15Z}. The
silicon velocity then decreased to 10,200\,km\,$\rm s^{-1}$ at
maximum light, at which point the velocity became relatively stable,
decreasing to only 9,700\,km\,$\rm s^{-1}$ at $+20.8$\,d. The pEWs of
\Siii\ increased slightly from our first epoch until $t \approx 5$\,d
and then faded away at later times.  The \Siiitmpsec\ and \Siiitmp\
features generally follow a similar evolution to that of \Siii, but
with lower velocities and weaker absorption.

The \Caii\ NIR triplet can be clearly identified in most of our
spectra, even extending to very late times.  The velocity of the
\Caii\ NIR PVF and \Siii\ have similar evolution, but the \Caii\
NIR HVF has much larger velocities than the \Caii\ NIR PVF.  For the
\Caii\ NIR HVF and \Caii\ NIR PVF, we measure a velocity of
25,130\,km\,$\rm s^{-1}$ and 15,955\,km\,$\rm s^{-1}$,
respectively, at $-16$\,d.  However, the velocity of the \Caii\ NIR
HVF dramatically decreases to \about 20,000\,km\,$\rm s^{-1}$ after
maximum brightness.  The velocity evolution of \Caii\ NIR line is 
consistent to that of normal-velocity 
(NV; \Siii\ velocity $<11,800$\,km\,$\rm s^{-1}$ at maximum brightness) 
SNe Ia studied in \citet{2015arXiv150207278S}. It is also evident from
the pEW measurements that the \Caii\ NIR HVF is stronger than the 
\Caii\ NIR PVF at early times. We find that the ratio of pEW(HVF) to pEW(PVF) 
is \about 5.3 at $-16$\,d, quickly decreases to unity at $t
\approx -5$\,d, and is well below unity at later times. 

\begin{figure}
	\centering
	\begin{tabular}{c}
		\includegraphics*[scale=0.75]{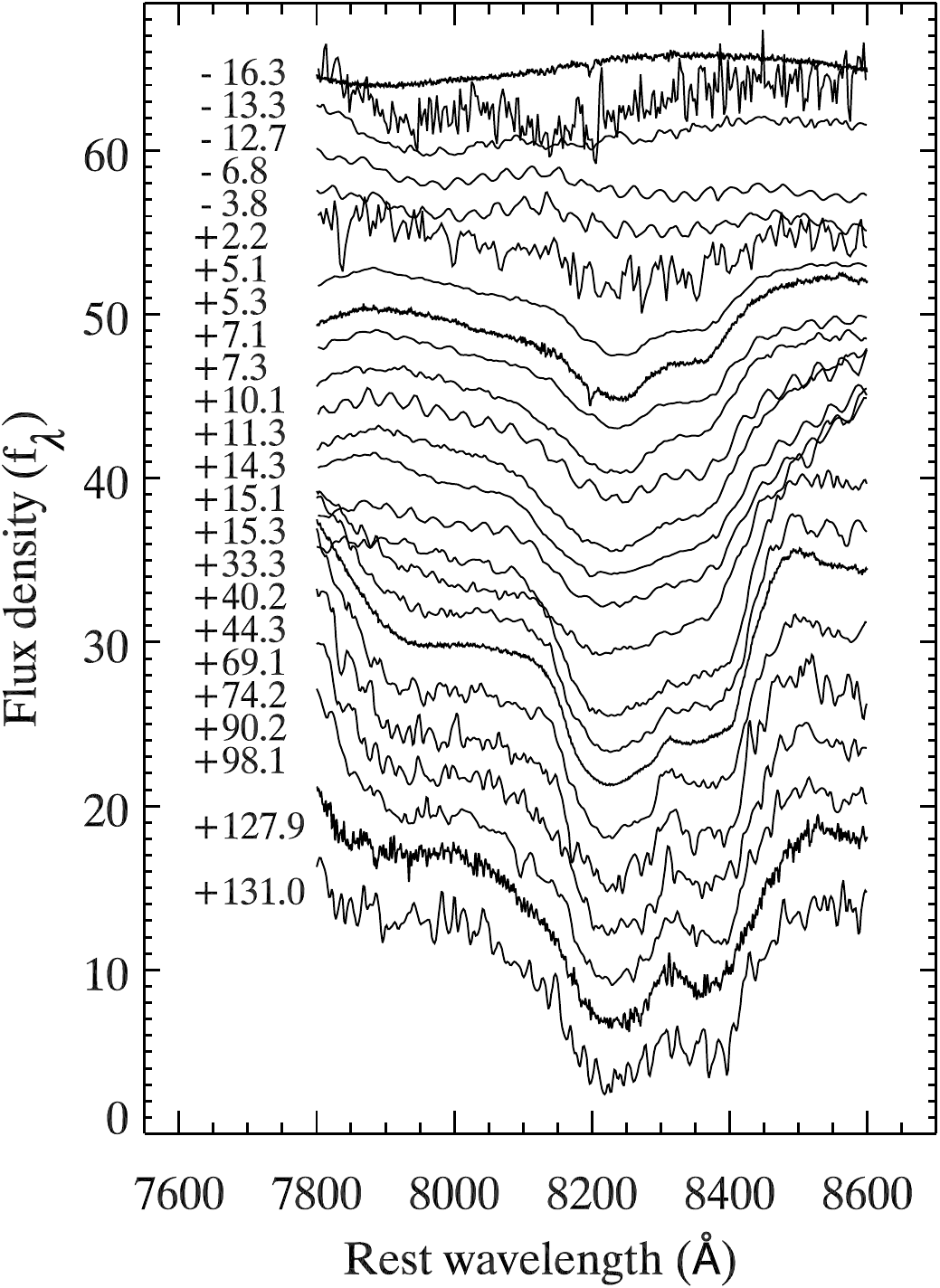}
	\end{tabular}
               \caption{The evolution of the \Caii\ NIR triplet with phase.
               }
        \label{cair_evolution}
\end{figure}

We also found that the \Caii\ NIR absorption showed increasingly
stronger absorption after maximum brightness. We display the evolution 
of the \Caii\ NIR triplet in Figure~\ref{cair_evolution}, which clearly shows the
strengthening NIR absorption with phase.  The absorption is still
clearly detected at $t = 131$\,d. 

\subsection{High-Resolution Spectroscopy}
\label{sec:hires}
\begin{figure*}
	\centering
	\begin{tabular}{c}
		\includegraphics*[scale=0.5]{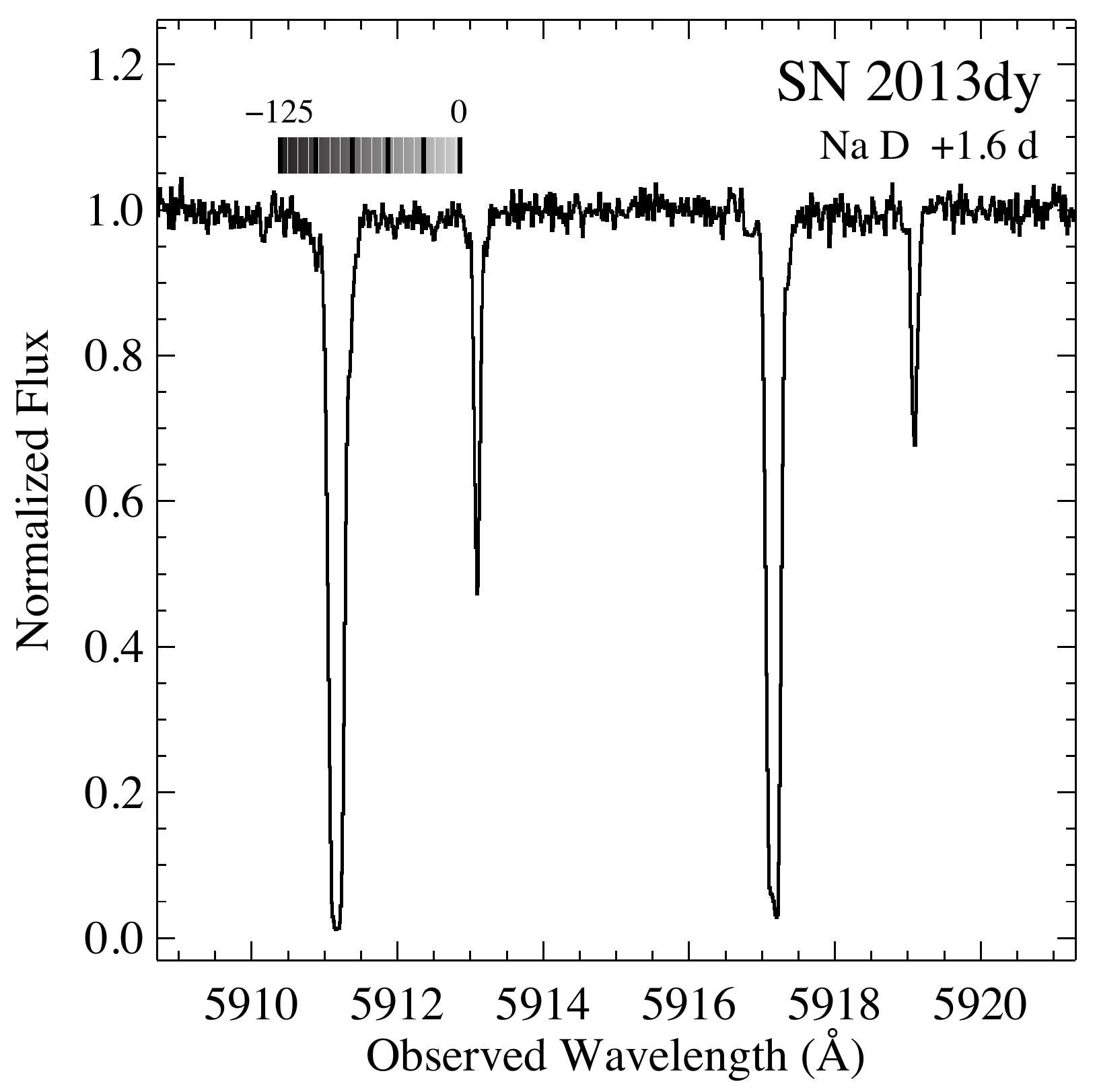}
	    \includegraphics*[scale=0.5]{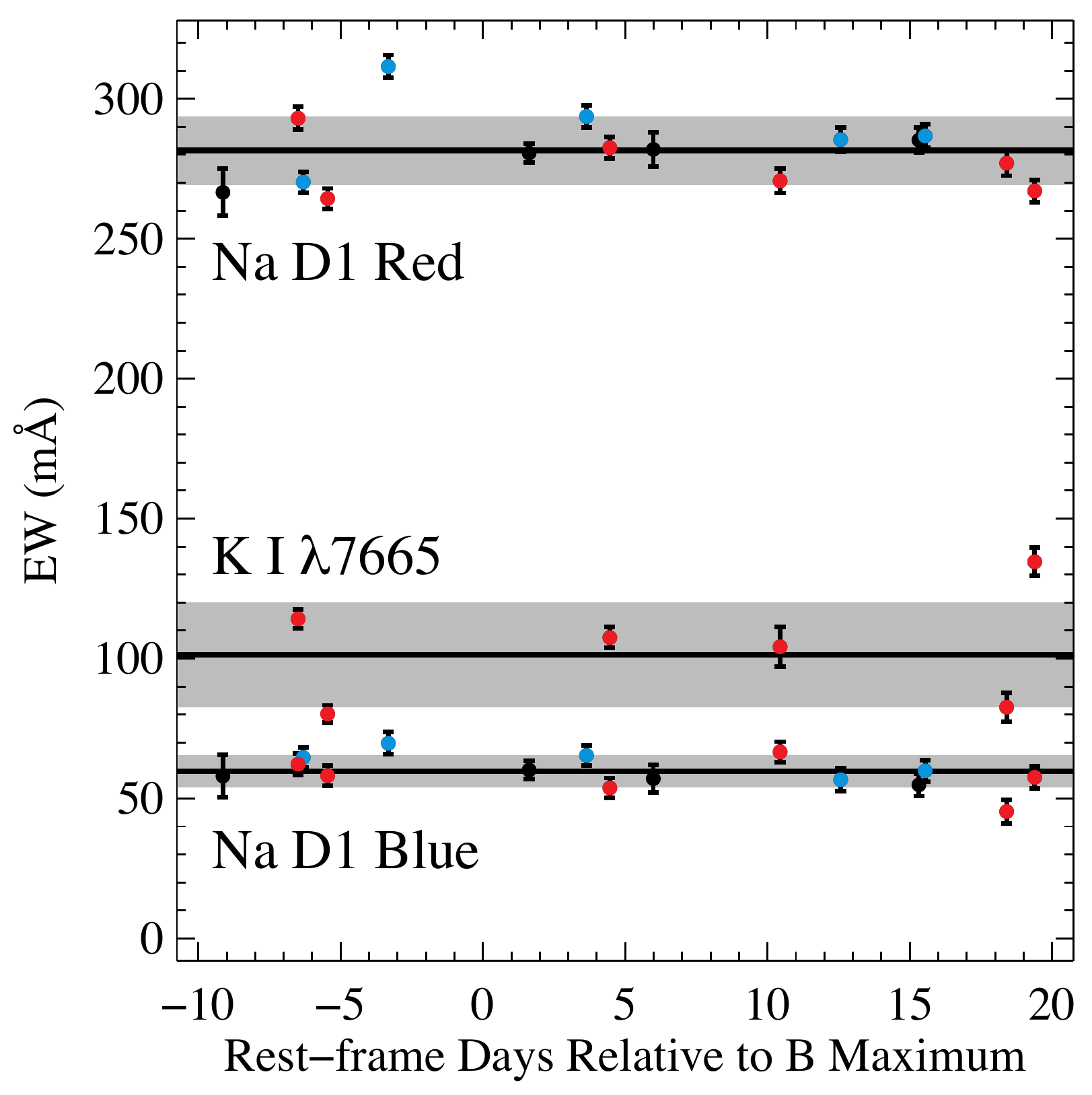}
	\end{tabular}
	\caption{Left: The Na~D absorption in the spectrum taken with VLT/UVES.
	A bar showing the velocity scale is presented on top of the spectrum.
	Right: The EWs as a function of phase of 
	the blue and red components for the Na~D1 line as well as the 
	\ion{K}{1} $\lambda 7665$ line. The spectra taken with different instruments
	are shown in different colors: HET/HRS (blue), Mercator/HERMES (red), and the
	rest of the instruments listed in Table~\ref{hres-log} (black).
    }
        \label{sn2013dy_hires}
\end{figure*}

Studies of the circumstellar environment of SNe~Ia are critical in constraining the 
progenitor models. By using a series of high-resolution optical spectra, 
recent studies have revealed that at least some SNe Ia have time-varying
narrow absorption lines \citep[e.g.,][]{2007A&A...474..931P,2009ApJ...702.1157S,
2012Sci...337..942D}. Some explanations suggest that these time-varying absorptions
could be caused by the interaction between SN ejecta and the circumstellar medium (CSM), 
which likely originates as gas outflows from the progenitor system.

In this work, we examined the high-resolution spectra at wavelengths corresponding
to absorption from Ca~H\&K, Na~D $\lambda\lambda 5890$, 5896, and
\ion{K}{1} $\lambda\lambda 7665$, 7699 at roughly the recession
velocity of the host galaxy.  We also examined absorption from the
5780~\AA\ diffuse interstellar band (DIB).  The recession velocity of
NGC~7250 is sufficiently large ($cz = 1166\rm\,km\,s^{-1}$) that there
should be no confusion between interstellar and/or circumstellar gas
in NGC~7250 and gas in the Milky Way.

Figure~\ref{sn2013dy_hires} (left) displays the Na~D absorption
in the $+1.6$~day spectrum of SN~2013dy.  There are two major
absorption features at roughly $-85$ and $+10\rm\,km\,s^{-1}$, with the
bluer component being stronger and saturated.  From the \ion{K}{1}
$\lambda 7665$ line, which is not saturated, we determine that the
$-85\rm\,km\,s^{-1}$ component is composed of two separate velocity
components at $-83.7$ and $-88.4\rm\,km\,s^{-1}$, with the redder
component being stronger.  Using the \citet{2011Sci...333..856S}
classification scheme, where the profile is classified by the velocity
of absorption features relative to the strongest Na absorption,
SN~2013dy has a ``Redshifted'' profile; however, the bluest
($-88.4\rm\,km\,s^{-1}$) component as seen in \ion{K}{1} is weaker
than the strongest absorption ($-83.7\rm\,km\,s^{-1}$), making the
classification somewhat complicated.  Using the
\citet{2013MNRAS.436..222M} classification scheme, where the profile
is classified by the velocity of all features relative to the systemic
velocity, SN~2013dy has a ``Blueshifted'' profile.

\begin{figure*}
	\centering
	\begin{tabular}{c}
		\includegraphics*[scale=0.75]{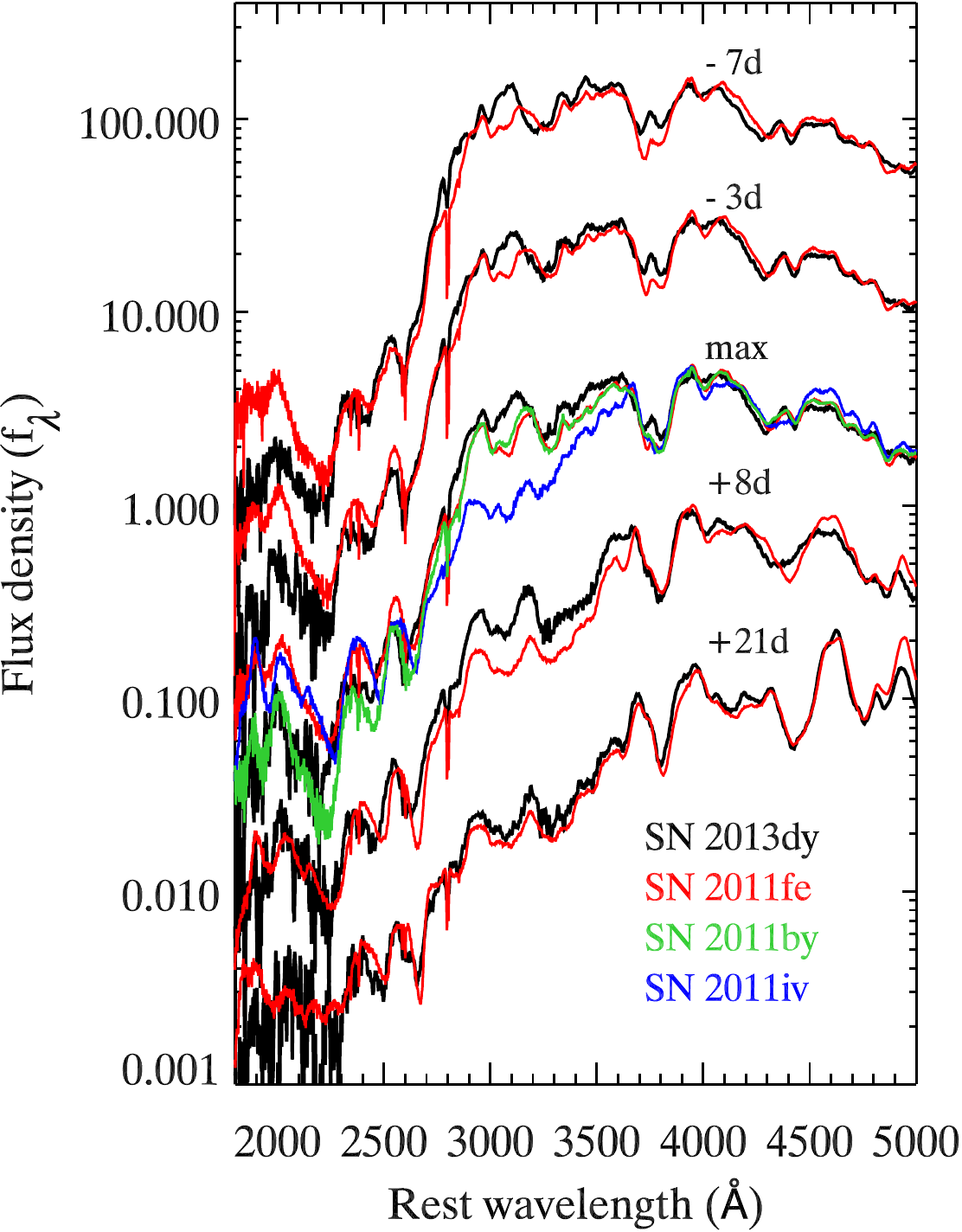}
		\includegraphics*[scale=0.76]{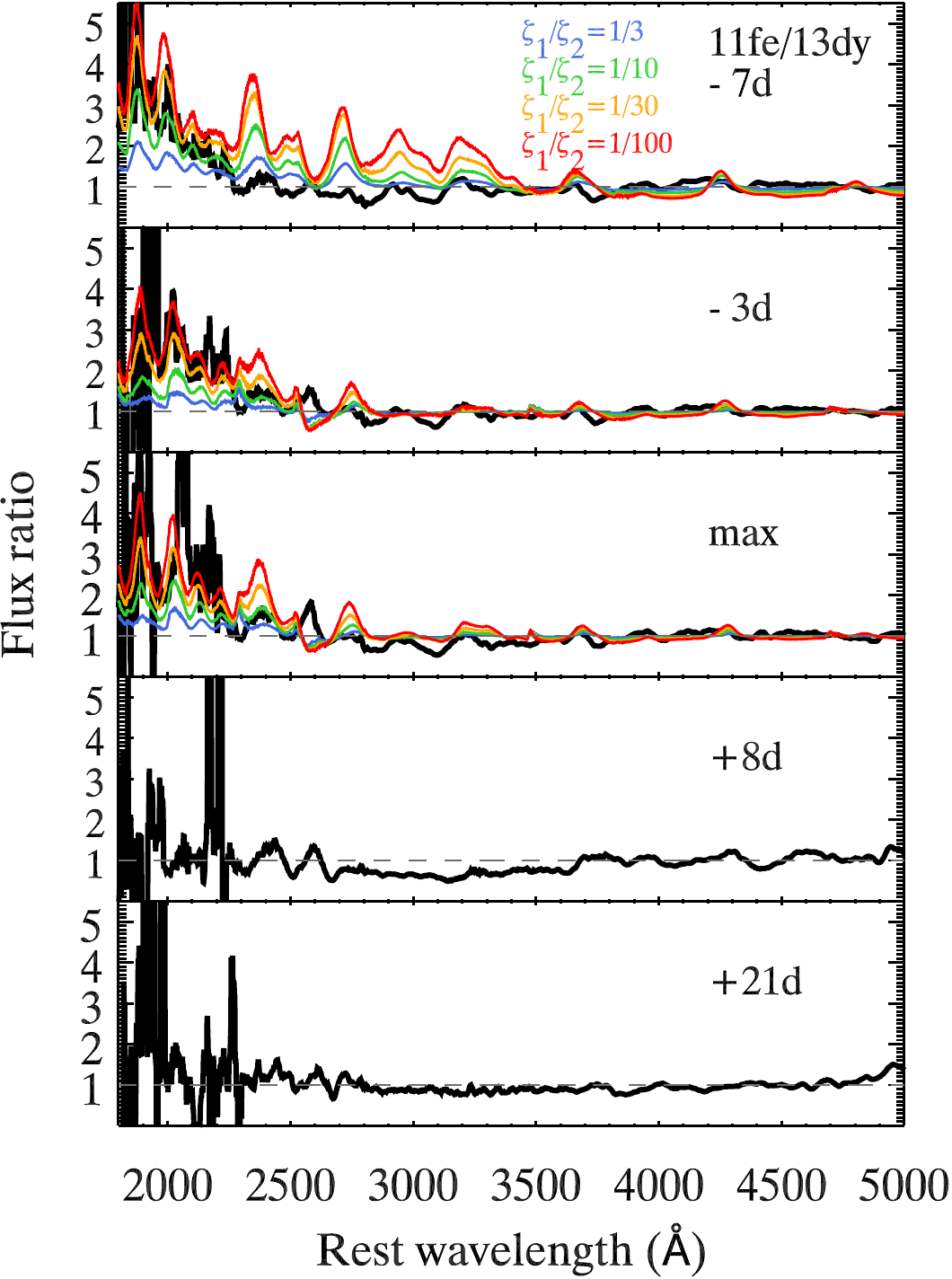}
	\end{tabular}
               \caption{Left: The comparison of SN~2013dy (black) and SN~2011fe (red) spectra 
               under $\rm5000\,\AA$. The maximum-light spectra of SN~2011by (green) and SN~2011iv (blue)
               are also compared. All spectra are dereddened
               and normalized accordingly (the normalized region is 3500--5000\,\AA).
               Right: The flux-ratio spectra (11fe/13dy) are shown in black.
               The predicted flux-ratio spectra with differences in metallicity factor $\zeta$
               are overplotted. Here the models from Days 7, 10, and 15 of \citet{2000ApJ...530..966L}
               are compared to the dataset at $t=-7$, $-3$, and 0 days relative to the $B$-band maximum brightness, respectively.
               The grey dashed line represents the line of equality.
               }
        \label{uv-compare}
\end{figure*}

Since the resolutions for the various instruments differ
significantly, we focus primarily on the equivalent widths (EWs) of
the absorption features.  However, in the cases where we have multiple
epochs from the same instrument, we also examined the line profiles in
detail.  Evaluating each subset from the same instrument also reduces
cross-instrument systematics.  There is no obvious change for any line
in our data.

We measure the EW of the blue and red components of the D1 and D2
lines separately.  We do not fit the features, but rather simply sum
the data over a wavelength range covering the features in all spectra.
This avoids the complication that the higher-resolution spectra have
clear flat-bottomed profiles, while lower-resolution spectra do not.
This flat-bottomed feature has two velocity components (as
determined by the \ion{K}{1} feature) offset by $<5\rm\,km\,s^{-1}$,
and at lower resolution they may not be completely resolved.  We also
measure the EW of the entire profile (for D1 and D2 separately) over a
wavelength range covering the entire feature.  These measurements
should be sensitive to any changes in absorption at velocities
different from that of the strong absorption features.

The statistical uncertainties were determined from the
root-mean-square scatter near the Na lines
\citep[e.g.,][]{2001PASP..113..920L}.  There are additional systematic
uncertainties related to the data reduction (and particularly the
telluric correction) as well as the continuum determination.  We
measure the latter uncertainty by varying the order and spacing of
breakpoints when fitting the B-splines to the continuum, measuring the
EW, and setting the standard deviation of the measurements to be the
uncertainty.  We determine the former uncertainty by comparing our
measurements taken with different instruments on the same night (with
the assumption that lines should not vary significantly in the span of
a few hours).  We present our measurements of the EWs for the blue and
red components for the Na D1 line as well as the \ion{K}{1} $\lambda
7665$ line in Figure~\ref{sn2013dy_hires} (right).

There is no evidence for variable absorption from any of the narrow
absorption features in our data.  Although the scatter for a given
measurement is significantly larger than the uncertainties in the
measurements (with $\chi^{2}_{\nu} \approx 10$), there is no coherent
change with phase.  Instead, we believe that a systematic bias affects
the HET/HRS observations, which have systematically high EWs for all
components of Na~D.  While we do not attempt to correct this potential
bias in the measurements, we note that the scatter is typically
$<10$\% of the measurement, indicating that the potential bias is
likely small. The peak-to-peak variations up to $\sim40\rm\,m\AA$ 
are also expected owing to the fractal patchy structure of the ISM \citep{2010A&A...514A..78P}.

In our highest S/N spectra, we detect the 5780~\AA\ DIB feature,
which is consistent with a single component having a velocity of $-20
\pm 10\rm\,km\,s^{-1}$.  This component is marginally consistent with
being at the same velocity as the redder $+10\rm\,km\,s^{-1}$ Na~D
absorption component.  The DIB has an EW of $60 \pm 10$~m\AA.  Using
the \citet{2013ApJ...779...38P} relation, this corresponds to $A_{V} =
0.3 \pm 0.2$~mag.  This estimate is comparable to, though lower
than, the extinction measurement determined from the SN~2013dy
photometry ($A_{V} = 0.64$~mag; see \S~\ref{sec:lc-fit}).

\subsection{UV Spectroscopy}
\label{sec:uv-spec}

We obtained 10 epochs of {\it HST}/STIS spectra in this work. The
multi-epoch UV spectra enable an investigation of the UV spectral
evolution at wavelengths as short as 1600\,\AA.  In
Figure~\ref{uv-compare} we compare the {\it HST} spectra of SN~2013dy
with those of SN~2011fe at similar epochs.  We also compare to
SNe~2011by and 2011iv at maximum brightness. SNe~2011by and 2011fe
have nearly identical optical colors, light-curve shapes, and spectra,
but have different UV continua \citep{2013ApJ...769L...1F,
  2015MNRAS.446.2073G}.  SN~2011iv is a spectroscopically normal
SN~Ia, but presents a relatively fast decline rate \cite[$\Delta
m_{15}(B) \approx 1.7$\,mag;][]{2012ApJ...753L...5F}.  All spectra were
dereddened and normalized to have the same mean flux between 3500 and
5000\,\AA.

At all epochs except for $t = +21$~d, SN~2013dy has excess flux at
$2800 < \lambda < 4000$~\AA\ relative to SN~2011fe.  Examining this
wavelength range at maximum brightness, we find a trend between the
amount of flux in this region (relative to the rest of the spectrum)
and light-curve shape.  Specifically, the continuum in this region is
strongest for SN~2013dy ($\Delta m_{15}(B) = 0.92$\,mag), followed by
SNe~2011by and 2011fe ($\Delta m_{15}(B) \approx 1.1$\,mag), and
finally SN~2011iv ($\Delta m_{15}(B) \approx 1.7$\,mag).  Therefore,
the amount of flux in this region may be linked to the amount of
\Nifs\ generated in the explosion. We also note that the
maximum-light spectra of SNe~2011by and 2013dy are relatively similar
for $\lambda < 2300$~\AA, with both having relatively less flux than
SN~2011fe at these wavelengths.  SN~2011fe also has excess flux at
these wavelengths, relative to SN~2013dy, at earlier epochs.

We further examine the UV evolution of SN~2013dy using the flux-ratio
spectra ($\rm f_{13dy}/f_{11fe}$) shown in the right-hand panel of
Figure~\ref{uv-compare}. Followed the method in
\citet{2013ApJ...769L...1F}, we overplot the flux-ratio spectra
predicted by the model of \citet{2000ApJ...530..966L} with differences
in metallicity factor $\zeta$. We find the SN~2011fe/SN~2013dy
flux-ratio spectra are better matched by the
\citet{2000ApJ...530..966L} flux-ratio spectra with $>$30 times
metallicity difference. This might suggest that SN~2013dy has a higher
progenitor metallicity than SN~2011fe; however, the other differences
between these SNe (light-curve shape and spectra; see
\S~\ref{sec:spec-class} for details) complicate this
possibility.

\begin{figure}
	\centering
	\begin{tabular}{c}
	\includegraphics*[scale=0.78]{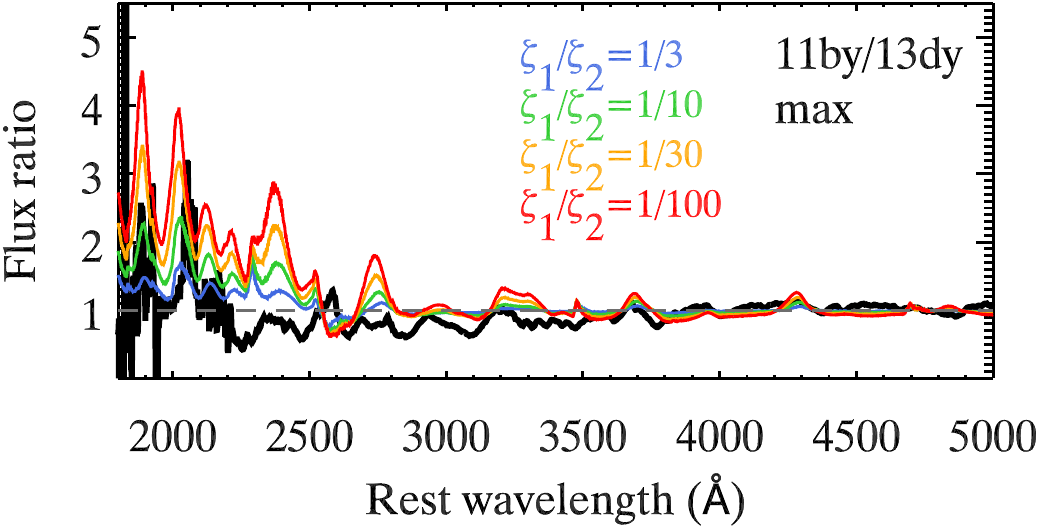}
    \end{tabular}
               \caption{Same as the right panel of Figure~\ref{uv-compare}, but
               using the flux-ratio spectrum of SN~2011by and SN~2013dy instead.
               }
        \label{uv-compare2}
\end{figure}

For comparison, we also determine the SN~2011by/SN~2013dy flux-ratio
spectrum at maximum light (see Figure~\ref{uv-compare2}). The result
shows that SN~2013dy and SN~2011by have similar continua in the UV.
This is consistent with the conclusion from
\citet{2013ApJ...769L...1F} and \citet{2015MNRAS.446.2073G} that
SN~2011by appears to have a higher progenitor metallicity than
SN~2011fe.

In addition to the difference in the continuum, the positions of the UV
spectral features of SN~2013dy are systematically bluer than those in
the spectra of SN~2011fe.  While lacking data at $\lambda < 2900$~\AA,
\citet{2012MNRAS.426.2359M} saw a similar trend for the features at
$2900 < \lambda < 3300$~\AA\ when comparing the mean spectra of SNe~Ia
with slower decline rates to those having faster decline rates.

\citet{2008ApJ...686..117F} defined the ``UV ratio'' as $R_{\rm UV} =
f_{\lambda}(2770\,{\rm \AA})/f_{\lambda}(2900\,{\rm \AA})$, finding that the 
UV ratio is strongly correlated with the luminosities of SNe~Ia. Brighter
SNe~Ia tend to have smaller $R_{\rm UV}$ than their fainter
counterparts.  For SNe~2013dy and 2011fe, we measure $R_{\rm UV} =
0.36$ and 0.42, consistent with their light-curve shapes.

The strong absorption feature at \about 3000\,\AA\ \citep[which is attributed
to \Feii;][]{1986ApJ...306L..21B} can be clearly identified in all UV spectra.  
However, the blending with many IGE lines in \Feii\ complicates the velocity measurement. 
Here we only measure the pEW of the feature as a whole at each epoch and present the result
in Figure~\ref{spec_evolution}. The pEW of \Feii\ increases slightly from \about 10\,\AA\ at
$-7$\,d to \about 60\,\AA\ at $+21$\,d. Our results compare well with
the evolution studied by \citet{2008ApJ...686..117F} using a larger
sample of SNe~Ia.

\subsection{Nebular Spectra}
\label{nebular}

\begin{figure}
	\centering
	\begin{tabular}{c}
		\includegraphics*[scale=0.75]{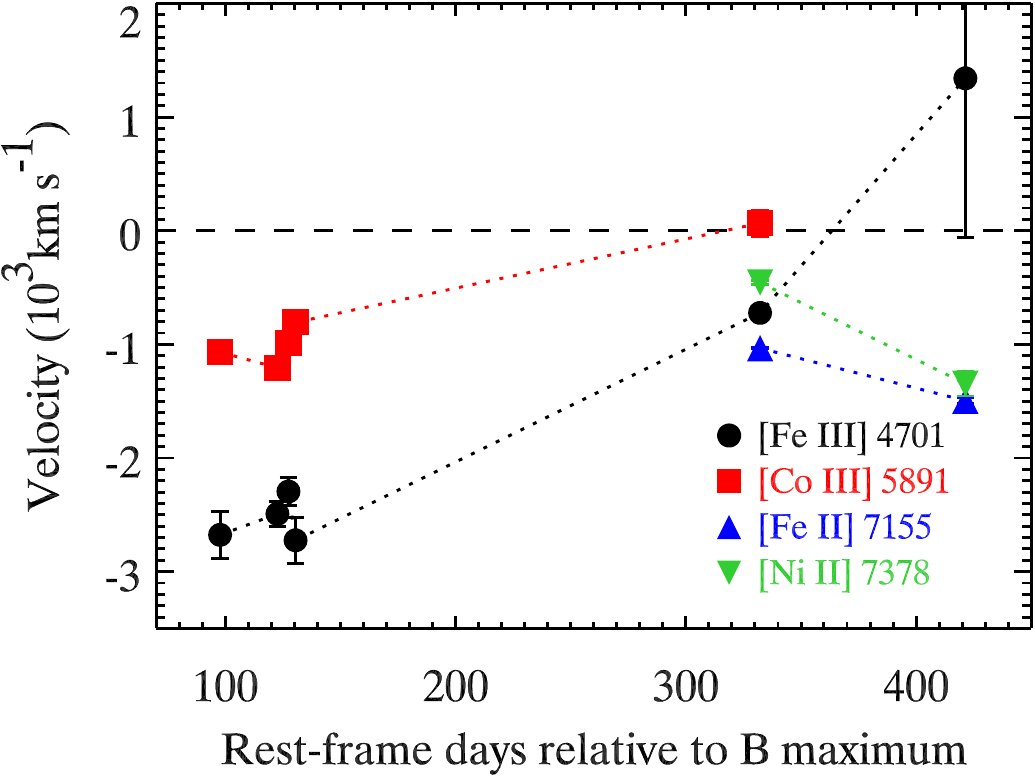}
	\end{tabular}
               \caption{The velocity determined from the emission lines 
               in SN~2013dy nebular spectra as a function of phase. 
               The velocity below/above the dashed line
               means the emission line is blueshifted/redshifted relative to the rest wavelength.
               }
        \label{nebv_evolution}
\end{figure}

\begin{figure}
	\centering
	\begin{tabular}{c}
		\includegraphics*[scale=0.75]{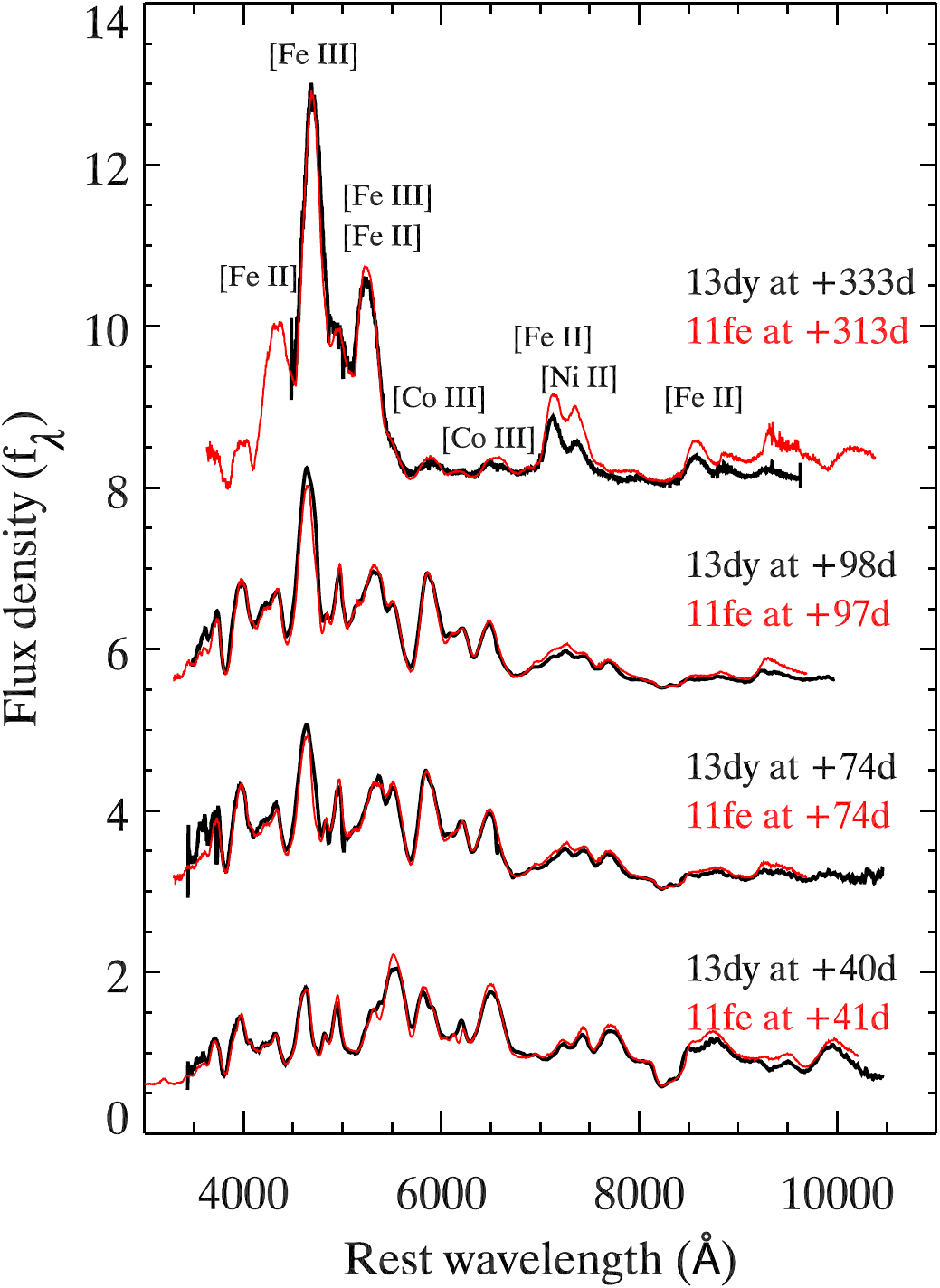}
	\end{tabular}
               \caption{Late-time spectra of SN~2013dy (black) and
               SN~2011fe (red). The spectra are dereddened and normalized 
               accordingly (the normalized region is 5000--6000\,\AA).
               }
        \label{nebular_compare}
\end{figure}

By \about 100\,d after maximum brightness, SNe~Ia enter the so-called
`nebular phase.' At this point, the SN becomes optically thin and
photons can escape from the very center of the ejecta.  Thus,
observations at these epochs are strongly constraining for SN~Ia
explosion models.  For these phases, the SN spectrum is dominated by
forbidden emission lines of IGEs (e.g., iron, nickel
and cobalt). In this work, we obtained 6 nebular spectra at $t >
100$\,d (three with $100 < t < 150$\,d with the remaining at $t >
300$\,d).

For each spectrum, we measured the velocity shift of the \Feiii,
\Feiineb, \Coiii\ and \Niii\ emission line features in the nebular
spectrum. Figure~\ref{nebv_evolution} 
shows the velocity of emission lines as a function of phase. \Feiii\ is 
one of the strongest spectral features in the nebular phase, and we are 
able to measure its velocity in most of the nebular spectra. It has a 
blushifted velocity of \about $-2600$~\kms\ relative to the rest wavelength 
of \Feiii\ at $t \approx 100$\,d. The velocity then decreases to $-720$~\kms\ at $t = 333$\,d,
and is redshifted to \about 1300~\kms\ at $t = 423$\,d. Similar trend
was also found in \citet{2013MNRAS.430.1030S} with a larger sample of SNe Ia.

\Coiii\ shows a similar evolution to \Feiii, but has a velocity that
is \about 1000~\kms\ closer to zero shift. \Feiineb\ and \Niii\ are
blended in wavelength space and can only be clearly decomposed in the
spectra at $t = 333$\,d and 423\,d. We deblend the line feature by
fitting the line with a double Gaussian profile. The velocities of
\Feiineb\ and \Niii\ are similar to that of \Feiii\ at $t = 333$\,d,
although an opposite evolution is found afterwards (the velocities of
\Feiineb\ and \Niii\ are further blueshifted at 423\,d).

In Figure~\ref{nebular_compare} we compare the late-time spectra of
SN~2013dy (starting \about 1 month after maximum brightness) with
similar phase spectra of SN~2011fe. The results show that both SNe are
very similar through $t \approx 100$~d.  However, there are noticeable
deviations for the \Feiineb\ and \Niii\ features as well as the
spectral region around 8500--9000\,\AA\ (dominated by
[$\mathrm{Fe}\,\textsc{ii}$] lines) at $t = 333$\,d. SN~2013dy appears
to have a relatively weaker $\Feiineb+\Niii$ feature than that of
SN~2011fe. At this late phase, the \Niii\ pEW of SN~2011fe ($1004 \pm
2$\,\AA) is about 1.6 times larger that of SN~2013dy ($630 \pm
6$\,\AA), but both SNe have similar \Feiineb\ pEWs ($739 \pm
8$\,\AA\ and $892 \pm 1$\,\AA\ for SN~2013dy and SN~2011fe,
respectively).

Following the diagnostic in \citet{2015MNRAS.446.2073G}, we determine
the relative strength of \Niii\ to \Feiineb, but using pEW instead of
the flux of the emission line.  Here the pseudo-continuum used to calculate the pEW
of emission line is defined with the local minima on each side of the line profile. 
We measure a pEW ratio of 0.85 and
1.12 for SN~2013dy and SN~2011fe, respectively.  Since the radioactive
\Nifs\ has a half-life of \about 6\,d, all of the nickel emission in
the nebular spectra at $t > 300$\,d is produced by stable \Nife.
However, the \Feiineb\ emission is produced by the combination of 
stable iron and radioactive \Nifs. This degeneracy complicates our
analysis. 

  For two identical SNe~Ia, except for the amount of (stable)
  $^{54}$Fe, radioactive $^{56}$Ni, and (stable) $^{58}$Ni generated
  in the explosion, their nebular spectra would be similar,
  with the largest discrepancy expected to be the flux ratio of
  \Niii/\Feiineb.  If the amount of stable iron-group elements is the
  same, but more $^{56}$Ni is generated, which will mostly decay to
  $^{56}$Fe at the time a nebular spectrum is observed, one would
  naively expect the \Niii/\Feiineb\ flux ratio to be smaller than in 
  an object with smaller $^{56}$Ni.  Similarly, if the amount of
  $^{56}$Ni is the same, but the amount of stable iron-group elements is
  smaller, then one would expect a smaller \Niii/\Feiineb\ flux
  ratio. 

  Since SN~2013dy has a smaller \Niii/\Feiineb\ flux ratio than
  that of SN~2011fe, SN~2013dy naively had a smaller amount of stable
  iron-group elements or a larger amount of $^{56}$Ni (or some
  combination) than did SN~2011fe.  For the latter case, and given the
  correlation between SN~Ia peak bolometric luminosity and the mass of
  \Nifs\ \citep{1979ApJ...230L..37A,1982ApJ...253..785A}, the
  Tully-Fisher distance to NGC~7250 (see \S~\ref{sec:bolometric}) may
  be underestimated.  Given the relative light-curve shapes for the
  two SNe, with SN~2013dy having a broader light curve than SN~2011fe,
  this is a reasonable explanation.  A separate distance measurement
  to the host galaxy (e.g., using Cepheid distance) is necessary to
  verify our results. 

  For the former case (SN~2013dy having a smaller amount of
  stable iron-group elements than SN~2011fe), the relative
  \Niii/\Feiineb\ flux ratio would suggest that SN~2013dy produced a
  smaller amount of stable to radioactive iron-group elements,
  consistent with SN~2013dy having a {\it smaller} metallicity than
  SN~2011fe, contrasting the results from our UV data (see
  Section~\ref{sec:uv-spec}).

  Again, the real situation may be more complicated, and some
  degeneracies need to be considered here.  SNe~Ia with higher
  metallicities not only produce more \Nife, but also more stable iron
  \citep{2003ApJ...590L..83T}.  That means the ratio
  \Niii/\Feiineb\ may not be necessarily larger with higher
  metallicity.  In addition, the prediction made by
  \citet{2003ApJ...590L..83T} is for the exact same explosion
  energies.  As we will discuss in \S~\ref{sec:spec-class}, this is
  unlikely to be the case for SN~2013dy and SN~2011fe.  Physical
  conditions other than progenitor metallicity could also contribute
  here \citep[e.g., central density;][]{2006MNRAS.368..187L,2012ApJ...757..175K,2013MNRAS.429.1156S}.
  \citet{2013MNRAS.429.1156S} showed that in Chandrasekhar-mass models
  the influence of central density (at ignition) on the ratio of \Niii/\Feiineb\ is even greater 
  than that of metallicity. They found that the models with higher central density generally have a
  higher ratio of \Niii/\Feiineb\ (assuming they have equal metallicity). It is
  currently difficult to disentangle the two situations for singular
  objects.  Nevertheless, the data appear to be more consistent with
  SN~2013dy having produced more \Nifs\ than SN~2011fe, with
  metallicity being less of an influence. 

\subsection{Spectroscopic Subclasses}
\label{sec:spec-class}
\begin{figure*}
	\centering
	\begin{tabular}{c}
		\includegraphics*[scale=0.75]{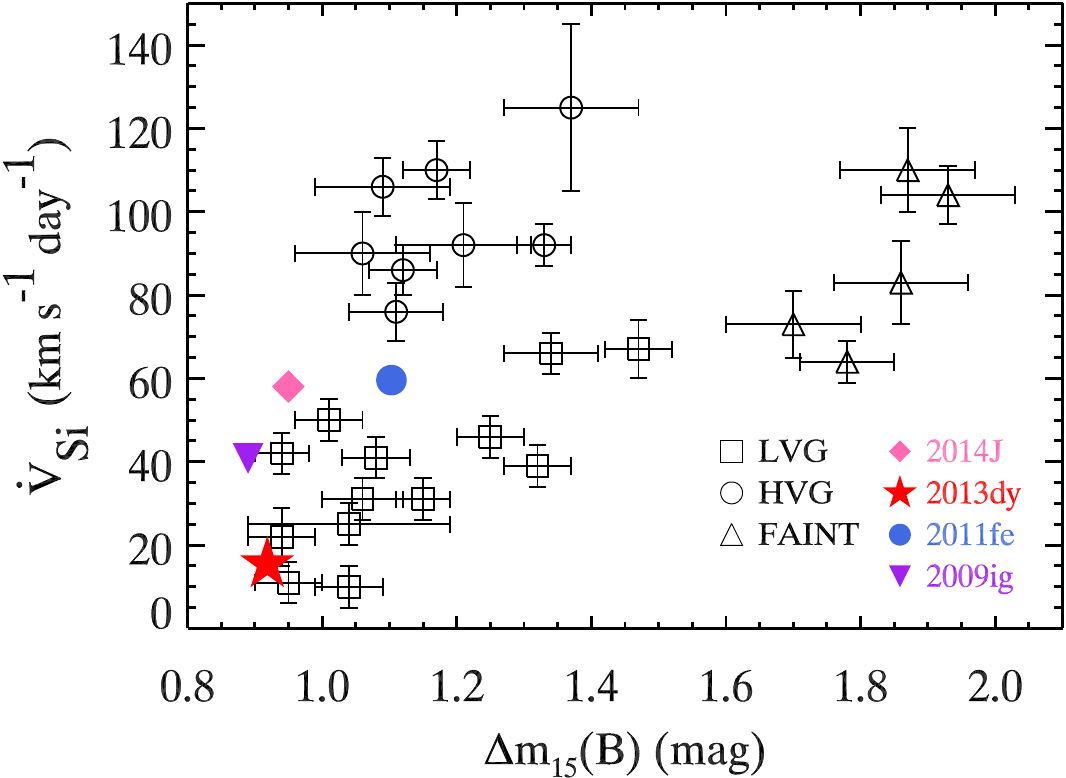}
		\includegraphics*[scale=0.75]{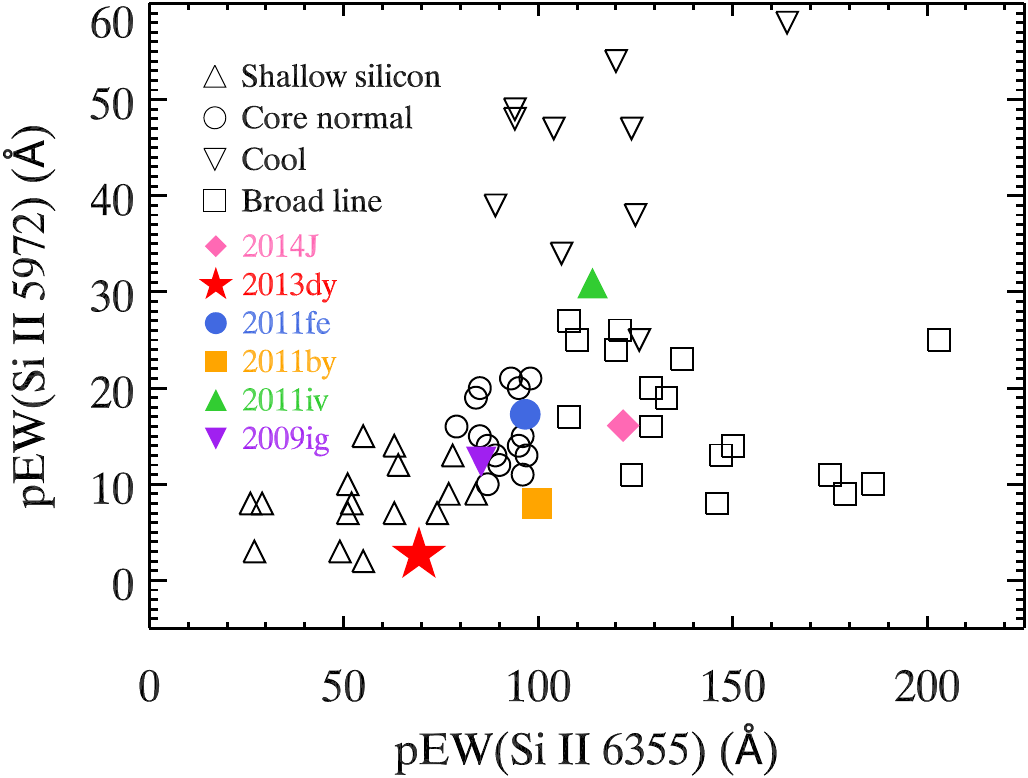}
	\end{tabular}
               \caption{Left: The gradient of \Siii\ velocity 
               ($\rm km\,s^{-1}\,day^{-1}$) as the
               function of $B$-band decline rate \deltam($B$). The sample and subclasses studied in
               \citet{2005ApJ...623.1011B} is overplotted.
               The sample is split into three different subclasses: 
               `low-velocity gradient' (LVG; open squares), `high-velocity gradient'
               (HVG; open circles) and `FAINT' (open triangles).
               Right: \Siiitmp\ pEW as the function of \Siii\ pEW.
               The sample and subclasses studied in \citet{2009PASP..121..238B}
               is overplotted. The sample is split into four
               different subclasses: `shallow silicon' (open triangles),
               `core normal' (open circles), `cool' (open downward triangles)
               and `broad line' (open squares).
               The position of SN~2013dy is represented by the red filled star.
               We also mark the positions of a variety of SNe~Ia for comparison.
               }
        \label{spec_subclass}
\end{figure*}

Previous studies which examined large sample of SNe~Ia and their
spectra at maximum light demonstrated the existence of several
subclasses of SNe~Ia \citep{2005ApJ...623.1011B, 2006PASP..118..560B}.
In Figure~\ref{spec_subclass} (left) we show the subclasses of
SNe~Ia as determined by the gradient of \Siii\ velocity and $B$-band
decline rate \deltam($B$) \citep{2005ApJ...623.1011B}. The SN~Ia
sample in \citet{2005ApJ...623.1011B} is split into three groups: a
`low-velocity gradient' (LVG) group, which have low \Siii\ velocity
gradients and natrually have lower values of \deltam($B$); a
`high-velocity gradient' (HVG) group, which have high \Siii\ velocity
gradients and low \deltam($B$); and a `FAINT' group including
SN~1991bg-like events that have higher values of \deltam($B$) and
naturally have larger velocity gradients.  SN~2013dy shows a low
\Siii\ velocity gradient ($\rm-15\,km\,s^{-1}\,day^{-1}$; measured
from maximum-light to $+21$\,d) and a slow decline rate ($\deltam(B) =
0.92$\,mag). Using these spectral indicators, we can classify SN~2013dy
as a LVG SN~Ia.

Figure~\ref{spec_subclass} (right) shows the subclasses of
SNe~Ia according to the pEW of \Siii\ and \Siiitmp\
\citep{2006PASP..118..560B}. \citet{2006PASP..118..560B} split the
SN~Ia sample into four groups: A `shallow-silicon' group, which have
low pEWs for both \Siii\ and \Siiitmp\ lines (and include
SN~1991T-like objects); a `core-normal' group, which have homogeneous
and intermediate pEWs; a `cool' group \citep[similar to the FAINT
group in][]{2005ApJ...623.1011B} with strong \Siiitmp\ lines relative
to \Siii; and a `broad-line' group, which present strong \Siii\
absorptions (and have much overlap with the HVG group). In this
scheme, we classify SN~2013dy as a `shallow-silicon' SN~Ia
corresponding to its relatively weak \Siii\ and \Siiitmp\ absorptions.

In addition to the samples of \citet{2005ApJ...623.1011B} and
\citet{2006PASP..118..560B}, we specifically added SNe~2009ig, 2011by,
2011fe, 2011iv, and 2014J to Figure~\ref{spec_subclass}.  All of these
SNe have maximum-light UV spectra \citep{2012ApJ...744...38F, 
2012ApJ...753L...5F,2013ApJ...769L...1F, 2013MNRAS.435..273F, 
2014MNRAS.443.2887F, 2014MNRAS.439.1959M}.  While
SNe~2009ig and 2014J have similar decline rates as SN~2013dy
($\deltam(B)=0.89$ and 0.95\,mag, respectively), both have larger
\Siii\ velocity gradients, with SN~2014J having a velocity gradient
somewhat intermediate to the LVG and HVG classes.  Nonetheless, all
three SNe are classified as LVG objects.  SN~2011fe, on the other
hand, is intermediate to the LVG and HVG groups.

Using purely spectral parameters, we classify SNe~2011by and 2011fe as
Core-normal, SN~2011iv as Cool, SN~2014J as Broad-lined, and SN~2009ig
as intermediate between Shallow-silicon and Core-normal.  This
indicates that there is no perfect match for SN~2013dy among the
SNe~Ia with maximum-light UV spectra.  However, from
Figure~\ref{spec_subclass}, we see that we are consistently filling
out the parameter space of SNe~Ia with objects that have UV spectra.
To this end, SN~2013dy represents an extreme object in velocity
gradient, \deltam($B$), pEW(\Siii), and pEW(\Siiitmp).

\section{Comparison to Models}
\label{sec:model}

With an excellent UV through NIR spectral sequence and a well-sampled
bolometric light curve, SN~2013dy is an ideal object to which one can
compare models.  Through these comparisons, we can constrain the
progenitor and explosion models of SN~2013dy.  For this purpose, we
select two primary models: The one-dimensional deflagration model W7
\citep{1984ApJ...286..644N} and the three-dimensional
delayed-detonation N5 model \citep{2013MNRAS.429.1156S}.  
These models are chosen based on their similar bolometric light-curve shapes 
to that of SN~2013dy \citep[e.g., see][]{2013MNRAS.436..333S}.
Spectral sequences were generated for these models using the radiative transfer
codes developed by \citet[ARTIS;][]{2009MNRAS.398.1809K}. The corresponding
model spectra were described in \citet{2009MNRAS.398.1809K} for the W7 model
and \citet{2013MNRAS.436..333S} for the N5 model.  In addition
to the nominal W7 model, we modify the W7 abundance profile to match
solar metallicity \citep{2009ARA&A..47..481A} in the outer layers
\citep[W7 $\rm Z_{\sun}$; see also][]{2012ApJ...753L...5F}.  This
model is useful for determining if the progenitor star had nonzero
metallicity.

\begin{figure}
	\centering
	\begin{tabular}{c}
		\includegraphics*[scale=0.75]{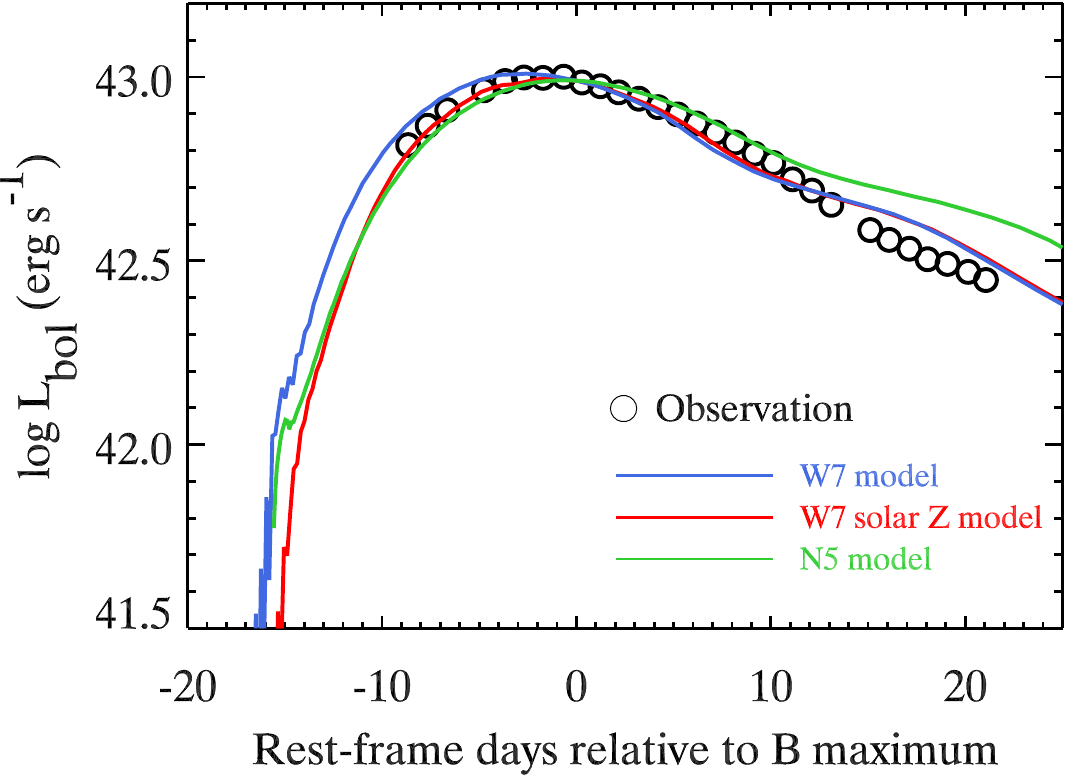}
	\end{tabular}
	\caption{The UVOIR bolometric light curve of SN~2013dy (same as Figure~\ref{bolometric-lc}) 
	is shown in open circles. We compare the data to model light curves generated 
	from W7 model (blue curve), solar-metallicity polluted W7 $\rm Z_{\sun}$ model (red curve) 
	and the three-dimensional delayed-detonation N5 model (green curve), respectively.
	Here the model light curve shown for N5 model is angle-averaged.
	The model light curves are shifted to match the bolometric flux of SN~2013dy at 
	the time of $B$-band maximum.
    }
    \label{model_compare_lc}
\end{figure}

In Figure~\ref{model_compare_lc} we compare the bolometric light curve
of SN~2013dy with the model bolometric light curves.  Because of the large
distance uncertainty of SN~2013dy, we shift the model light curves
such that their flux at the time of $B$-band maximum brightness match that of SN~2013dy.
By doing this, we are examining the shape of the bolometric light curve 
instead of the absolute luminosity.
Future observations may more accurately determine the peak luminosity of SN~2013dy
and thus better constrain the explosion models.

\begin{figure*}
	\centering
	\begin{tabular}{c}
		\includegraphics*[scale=0.88]{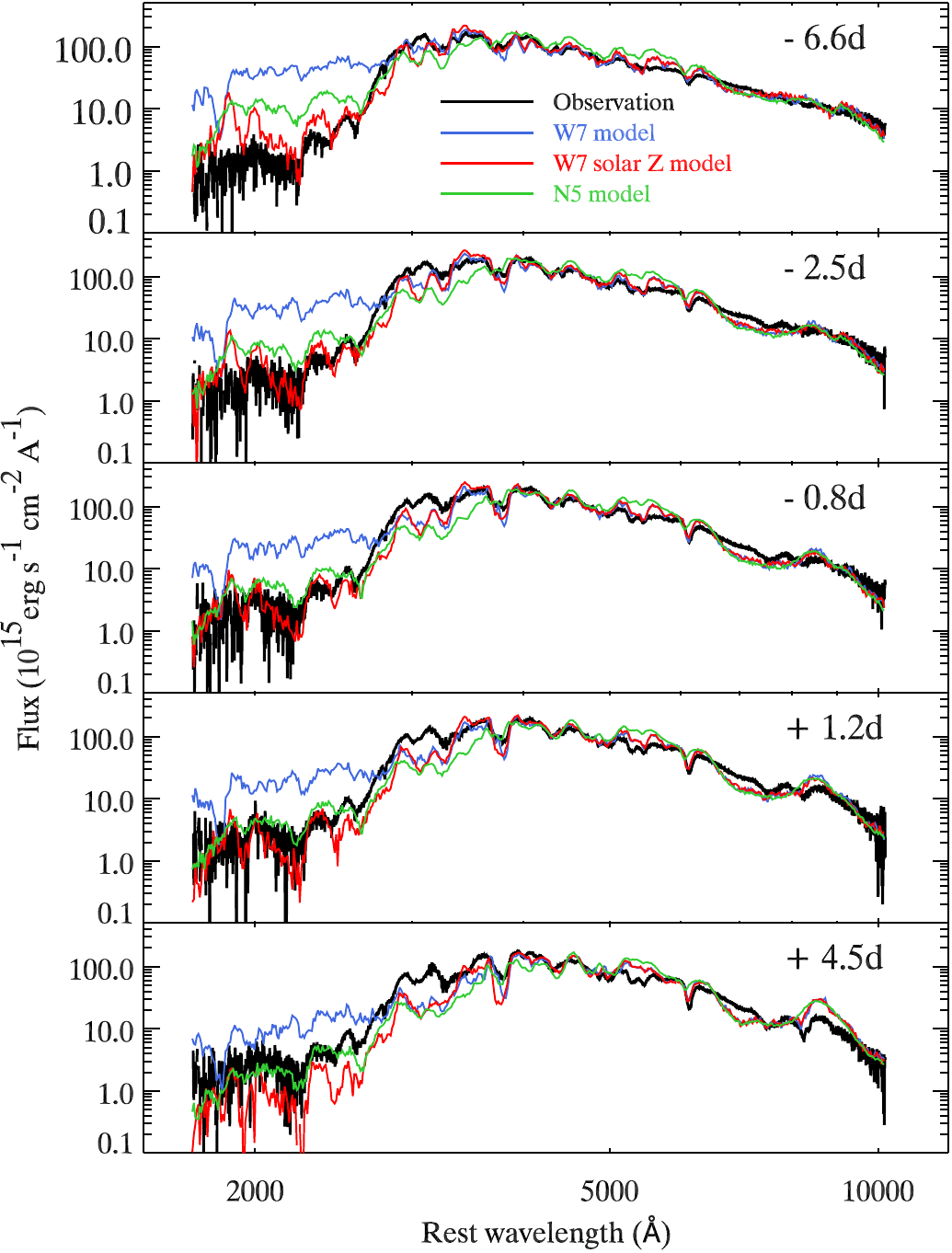}
	    \includegraphics*[scale=0.88]{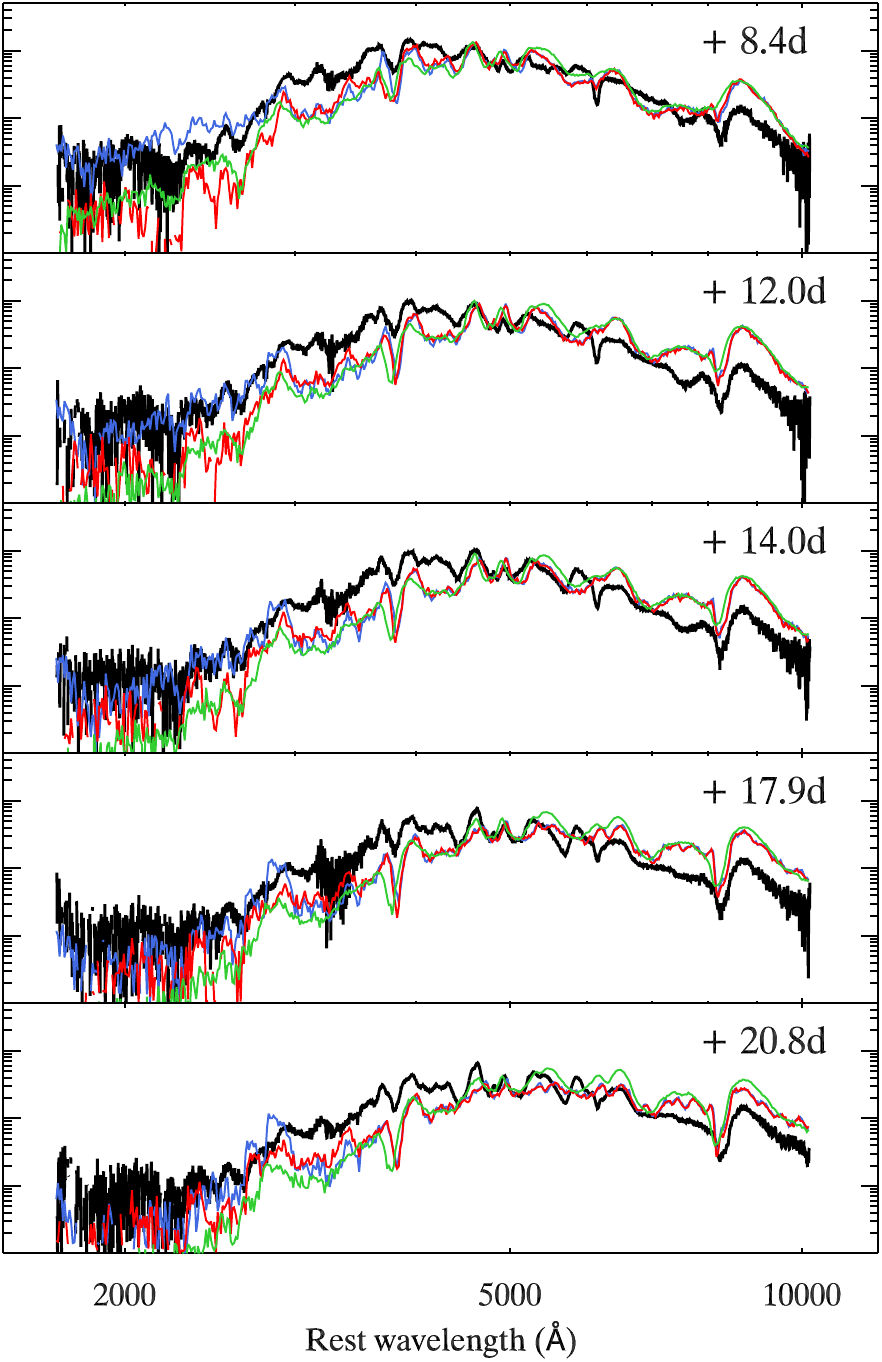}

	\end{tabular}
	\caption{The {\it HST} spectra at different epochs (black curves). We compare
	the data to model spectra generated from W7 model (blue curves),
	solar-metallicity polluted W7 $\rm Z_{\sun}$ model (red curves) and the three-dimensional
	delayed-detonation N5 model (green curves), respectively.
	The model spectra are scaled with the same multiplicative factor used to 
	shift the model bolometric light curve in Figure~\ref{model_compare_lc}.
             }
        \label{model_compare_spec}
\end{figure*}

At early phases ($t \leq 0$\,d), the model light curves generated
from W7 $\rm Z_{\sun}$ and N5 models are nearly identical, and both
are a better match to the data than the W7 model.  At later phases 
($t \gtrsim 0$\,d), the N5 light curves start to deviate from the data, 
having a slower decline, while the W7 and W7 $\rm Z_{\sun}$ evolve in a 
similar fashion as SN~2013dy.  The slower decline found in N5 model could 
be a reflection of the larger iron-group element mass compared to W7 model, which will 
result in a higher opacity and temperature.  The difference between the 
W7 and W7 $\rm Z_{\sun}$ models are negligible at later epochs since the W7 and 
W7 $\rm Z_{\sun}$ models only differ in the outer layers. The postmaximum
evolution of both models should resemble to each other as expected.  We further
demonstrate this effect in Figure~\ref{model_compare_spec} by
investigating the details of the spectral sequence.

Previous theoretical studies have shown that the UV SED is extremely
sensitive to the progenitor metallicity as it significantly alters the
UV opacity, which is dominated by line blanketing
\citep{1998ApJ...495..617H, 2000ApJ...530..966L}.  SNe~Ia with
identical optical spectra could have (and have shown) large differences
in the UV \citep{2013ApJ...769L...1F}.  In
Figure~\ref{model_compare_spec}, we compare the model spectra with
those of SN~2013dy.  To be consistent, the model spectra are scaled with 
the same multiplicative factor used to shift the model bolometric light curve
in Figure~\ref{model_compare_lc}. We exclusively examine the {\it HST} spectra,
since almost all of the discriminating power is in the UV.

For the phases earlier than a week after maximum brightness, the W7,
W7 $\rm Z_{\sun}$, and N5 models have similar optical spectra with all
generally agreeing with the data.  However, as already noticed by
\citet{2012ApJ...753L...5F}, the W7 and W7 $\rm Z_{\sun}$ models start
to have large differences at far-UV wavelengths ($\lambda \leq
2800$~\AA), which is exactly the region shown to be sensitive to the
progenitor metallicity \citep[e.g.,][]{2000ApJ...530..966L}.  While
the UV flux of the W7 model is significantly larger than that of
SN~2013dy ($\gtrsim$50 times at some wavelengths), the W7 $\rm
Z_{\sun}$ and N5 models are generally in agreement with the data.  
This result seems to be a conflict with our previous conclusion
since the N5 model does not have solar metallicity admixture like
the W7 $\rm Z_{\sun}$ model. However, in the three-dimensional delayed-detonation 
models, the IGE-rich deflagration ashes will rise due to the turbulent motions
\citep[e.g.,][]{2013MNRAS.436..333S}. This will lead to a mixture of burning 
ashes at higher velocities and cause the suppression of UV flux, which is similar to the effect of
progenitor metallicity in W7 $\rm Z_{\sun}$ model.
For our earliest epoch ($t = -6.6$~d), even the N5 model has more UV flux
than the data, but at this time, the W7 $\rm Z_{\sun}$ model is
relatively consistent with the observations.

For the phases later than a week after maximum brightness, the models
are less consistent with the observations.  These discrepancies might
be caused by systematic effects in the model spectra related to
approximations in the treatment of excitation and ionisation in our
radiative transfer simulations. Since full non-LTE calculations are currently unavailable for these explosion models, this limitation is expected to introduce some uncertainty to
the radiative transfer simulations in this work.

To summarize, we find that the W7 $\rm Z_{\sun}$ model is in good agreement
with SN~2013dy.  In terms of bolometric behavior and early-time UV
spectra, it performs significantly better than the zero-metallicity W7
model.  For the earliest epoch of spectroscopy and at later times, the
W7 $\rm Z_{\sun}$ model also performs slightly better than the
delayed-detonation N5 model. The UV data were critical in making 
this determination.

\section{Conclusions}
\label{sec:conclusion}
SN~2013dy was discovered only 2.4\,hr after explosion in NGC~7250,
which makes it the SN~Ia with the earliest known detection.  In this
work we present high-quality photometric and spectroscopic data. The
photometric observations were taken from $t = -16$ to $+337$\,d 
(relative to maximum brightness).
Ten epochs of UV through NIR spectra were taken with {\it HST}/STIS.
For phases of $t = -16$ to $+480$\, additional low- and
high-resolution optical spectra were obtained with a variety of
ground-based facilities.

Our photometric analysis shows that SN~2013dy has a relatively slow
decline rate ($\Delta m_{15}(B) = 0.92$\,mag) and is intrinsically bluer
than SN~2011fe.  Consistent with previous results, SN~2013dy, being a
slow-decling SN~Ia, has color curves that peak later and have
shallower later-time slopes than those of faster-declining SNe~Ia.  We
generated a UVOIR bolometric light curve for SN~2013dy (1600--18,000\,\AA) 
with both photometric and spectroscopic datasets, but large 
uncertainties in the distance to SN~2013dy prevented a
precise measurement of its peak bolometric luminosity.

We also determined the physical parameters of NGC~7250, the host
galaxy of SN~2013dy.  NGC~7250 is blue with evidence of strong
star-formation.  The relatively low \mstellar\ suggests that the
environment of SN~2013dy is likely metal poor. This seems to be a
conflict with the results suggested by the UV spectroscopy.  However,
using the global properties of host galaxy as proxy to constrain the
SN progenitor could present large uncertainties.  The host
  parameters determined in this work may not well reflect the real
  physical conditions of the SN birthplace.

In the earliest spectra (1.6 to 5\,d after the explosion), there are
strong \ion{C}{2} absorption features, indicating the evidence of
unburned material exists at the outer layers of the ejecta.
Using the classification schemes defined by
\citet{2005ApJ...623.1011B} and \citet{2009PASP..121..238B}, we
further classified SN~2013dy as a SN~Ia with low-velocity gradient
(LVG) and shallow-silicon absorption. Therefore, SN~2013dy is
spectroscopically different from Core-normal SNe similar to SN~2011fe.

SN~2013dy also has strong \Caii\ NIR HVF at early phases. The velocity
of the \Caii\ NIR HVF is \about 10,000\,km\,$\rm s^{-1}$ higher than
that of the \Caii\ NIR PVF. We find that the \Caii\ NIR HVF was
  stronger than the \Caii\ NIR PVF a few days after the
  explosion.  However, the \Caii\ NIR HVF quickly became weaker than
\Caii\ NIR PVF immediately after $t = -5$\,d. Although the velocity of
the \Caii\ NIR triplet decreased quickly with phase, we saw
increasingly stronger absorptions for the \Caii\ NIR triplet from
maximum brightness until $t = +131$\,d.

We obtained 15 epochs of high-resolution spectra, which is one of 
largest known single set of high-resolution spectra for a SN~Ia. Examining
narrow absorption features including \ion{Na}{1}~D, \ion{K}{1}
$\lambda 7665$, and the 5780~\AA\ DIB, we find no evidence of temporal
variability.

We compared the {\it HST} UV spectra to those of SNe~2011by, 2011fe,
and 2011iv at similar phases.  At maximum brightness, there appears to
be a correlation between the amount of UV flux at $2800 < \lambda <
4000$\,\AA\ and decline rate with SN~2013dy having both the slowest
decline rate and highest UV flux of the sample.  By examining the
flux-ratio spectra, we find that SN~2013dy has a depressed far-UV
continuum ($\lambda < 2300$\,\AA) relative to that of SN~2011fe,
consistent with predictions of a higher metallicity progenitor for
SN~2013dy.  However, differences in the two SNe complicate this
interpretation.  SN~2013dy also shows systematically blueshifted UV
features relative to SN~2011fe.

  The late-time spectra of SNe~2013dy and 2011fe are very
  similar from $t = 40$ to 100~d.  However, we find that SN~2013dy has a
  smaller ratio of \Niii/\Feiineb\ than SN~2011fe at $t \approx
  333$\,d. This may suggest that SN~2013dy synthesized more \Nifs\
  than SN~2011fe assuming both SNe generated the same amount of stable
  iron and \Nife, but we note that metallicity could have some
  effect here.

By comparing the bolometric light curve and spectral sequence of
SN~2013dy to that of SN~Ia explosion models, we find that the W7 $\rm
Z_{\sun}$ model (solar metallicity polluted W7 model) is in good agreement
with the entire SN~2013dy dataset.  The W7 $\rm Z_{\sun}$ model
is much more consistent than the zero-metallicity W7 model and
slightly more consistent than the delayed-detonation N5 model.
SN~2013dy again demonstrates that UV data are critical for
understanding the progenitor metallicity for SNe~Ia.

SN~2013dy is one of the best-studied SNe~Ia.  It is very nearby and
bright, and was discovered immediately after explosion.  Follow-up
observations were obtained consistently, spanning both a large time
frame and wavelength range. This exquisite dataset shows that a
detailed study of a single well-observed SN~Ia can provide unique
information about the progenitors and explosions of SNe~Ia.

\section*{acknowledgments} 
  Based on observations made with the NASA/ESA {\it Hubble Space
    Telescope}, obtained from the Data Archive at the Space Telescope
  Science Institute, which is operated by the Association of
  Universities for Research in Astronomy, Inc., under NASA contract
  NAS 5--26555. These observations are associated with program
  GO--13286.  We thank the Telescope Time Review Board for allowing
  these observations to begin before the official start of Cycle 21.
  We especially thank the STScI staff for accommodating our
  target-of-opportunity program.  A.\ Armstrong, R.\ Bohlin, S.\
  Holland, and D.\ Taylor were critical for the execution of this
  program. Some of the data presented herein were obtained at the W.M. Keck
  Observatory, which is operated as a scientific partnership among the
  California Institute of Technology, the University of California, and
  NASA; the observatory was made possible by the generous financial
  support of the W.M. Keck Foundation.

 This work was supported by the Deutsche Forschungsgemeinschaft 
  via the Transregional Collaborative Research Center TRR 33 ``The Dark 
  Universe'' and the Excellence Cluster EXC153 ``Origin and Structure 
  of the Universe''. A.S. is supported by the DFG cluster of excellence 
  Origin and Structure of the Universe. G.P. acknowledges support provided by the Millennium 
  Institute of Astrophysics (MAS) through grant IC120009 of the Programa 
  Iniciativa Cientifica Milenio del Ministerio de Economia, Fomento y 
  Turismo de Chile. J.V. is supported by Hungarian OTKA Grant NN 107637. 
  J.M.S. is supported by a National Science Foundation (NSF) 
  Astronomy and Astrophysics Postdoctoral 
  Fellowship under award AST-1302771. J.C.W.'s supernova group at UT Austin 
  is supported by NSF Grant AST 11-09801.
  The work of A.V.F's group at U.C. Berkeley was made possible by NSF grant AST-1211916, 
  the TABASGO Foundation, Gary and Cynthia Bengier, and the Christopher 
  R. Redlich Fund. S.T. acknowledges support by TRR 33 ``The Dark Universe''
  of the German Research Foundation (DFG).
  R.P.K. thanks the NSF for support through AST-1211196.
  M.D.S. gratefully acknowledges generous support provided by the Danish Agency for Science and 
  Technology and Innovation realized through a Sapere Aude Level 2 grant.
  This paper is partially based on observations made with ESO telescopes at the La Silla Paranal 
  Observatory under programme 091.D-0780(A).
    
  HET is a joint project of the University of Texas at Austin, 
  the Pennsylvania State University, Stanford University, Ludwig-Maximilians-Universitat 
  Munchen, and Georg-August-Universitat Gottingen. The HET is named in honor of 
  its principal benefactors, William P. Hobby and Robert E. Eberly. The 
  Marcario Low Resolution Spectrograph is named for Mike Marcario of High 
  Lonesome Optics who fabricated several optics for the instrument but died 
  before its completion. The LRS is a joint project of the HET partnership 
  and the Instituto de Astronomia de la Universidad Nacional Autonoma de Mexico.  
  We thank the staff at McDonald Observatory for their excellent work during the observations.
  RP acknowledges support by the European Research Council under ERC-StG grant 
  EXAGAL-308037 and by the Klaus Tschira Foundation.
  
  We thank the RATIR instrument team and the staff of the Observatorio Astrono\'mico Nacional on Sierra San Pedro Ma   \'rtir.  RATIR is a collaboration between the University of California, the Universidad Nacional Autono\'ma de Me\'xico, NASA Goddard Space Flight Center, and Arizona State University, benefiting from the loan of an H2RG detector from Teledyne Scientific and Imaging. RATIR, the automation of the Harold L. Johnson Telescope of the Observatorio Astrono\'mico Nacional on Sierra San Pedro Ma\'rtir, and the operation of both are funded by the partner institutions and through NASA grants NNX09AH71G, NNX09AT02G, NNX10AI27G, and NNX12AE66G, CONACyT grant INFR-2009-01-122785, UNAM PAPIIT grant IN113810, and a UC MEXUS-CONACyT grant.

\bibliographystyle{mn2e}
\bibliography{sn2013dy}
\label{lastpage}

\end{document}